\pdfoutput=1

\documentclass[12pt]{article}

\usepackage{amsfonts}
\usepackage{amsmath}
\usepackage{amssymb}
\usepackage{bigints}
\usepackage{booktabs}
\usepackage[nosort]{cite}
\usepackage{color}
\usepackage{dsfont}
\usepackage{float}
\usepackage{framed}
\usepackage{graphicx}
\usepackage{indentfirst}
\usepackage{mathrsfs}
\usepackage{multirow}
\usepackage{pdflscape}
\usepackage{setspace}
\usepackage{subdepth}
\usepackage{subfig}
\usepackage{titlesec}
\usepackage[dotinlabels]{titletoc}
\usepackage{wrapfig}
\usepackage[all]{xy}
\usepackage{young}
\usepackage[vcentermath]{youngtab}
\usepackage{relsize}
\usepackage{stackengine}
\usepackage{datetime}

\usepackage{hyperref}

\numberwithin{equation}{section}

\usepackage{verbatim}

 \newcommand{\reef}[1]{(\ref{#1})}

\newcommand{\be}{\begin{equation}}
\newcommand{\ee}{\end{equation}}

\newcommand{\bml}{\begin{multline}}
\newcommand{\emll}{\end{multline}}

\newcommand{\nn}{\nonumber}

\def\({\left(} \def\){\right)}
\def\[{\left[} \def\]{\right]}

\def\sd{\mathscr{D}}

\def\da{a^{\dagger}}

\def\al{\alpha}

\def\mO{\mathcal{O}}

\def\eps{\epsilon}

\def\v{\vec}

\def\a{\alpha}
\def\b{\bar}
\def\g{\gamma}

\def\lam{\lambda}

\def\d{\partial}

\def\o{\omega}

\newcommand{\la}{\langle}
\newcommand{\ra}{\rangle}

\newcommand{\bea}{\begin{eqnarray}}
\newcommand{\eea}{\end{eqnarray}}

\usepackage[left=2cm,right=2cm,top=2cm,bottom=2cm]{geometry}
\titleformat{\section}{\large\bfseries}{\thesection.}{4pt}{}
\titlespacing{\section}{0pt}{22pt}{6pt}

\titleformat{\subsection}{\normalfont\bfseries}{\thesubsection.}{4pt}{}
\titlespacing{\subsection}{0pt}{18pt}{6pt}

\titleformat{\subsubsection}{\normalfont\itshape}{\thesubsubsection.}{4pt}{}
\titlespacing{\subsubsection}{0pt}{16pt}{6pt}

\def\ie{\begin{equation}\begin{aligned}}
\def\fe{\end{aligned}\end{equation}}

\def\tilde{\widetilde}
\def\t{\tilde}

\def\bar{\overline}

\def\d{\partial}

\def\1{{\mathds 1}}

\def\mL{\mathcal{L}}

\def\o{\omega}
\def\v{\vec }

\DeclareFontShape{OT1}{cmr}{mx}{n}%
    {<->cmr10}{}
\newcommand{\mytitlefont}{\fontseries{mx}\selectfont}
\DeclareMathAlphabet{\titlemath}{OT1}{cmr}{mx}{n}

\newcommand{\bi}{\begin{itemize}}
\newcommand{\ei}{\end{itemize}}

\newcommand{\sss}{\subsubsection}

\usepackage{bm}

\usepackage{tikz}
\tikzset{every picture/.style={line width=0.75pt}} 

\begin{document}

\begin{titlepage}

\begin{center}

~\\[1cm]

{\fontsize{20pt}{0pt} \mytitlefont Loop diagrams in the kinetic theory of waves }\\[10pt]

~\\[0.2cm]

{\fontsize{14pt}{0pt}Vladimir Rosenhaus{\small $^{1}$}, Daniel Schubring{\small $^{1}$}, Md Shaikot Jahan Shuvo{\small $^{1}$}, and Michael Smolkin{\small $^{2}$}}

~\\[0.1cm]

\it{$^1$ Initiative for the Theoretical Sciences}\\ \it{ The Graduate Center, CUNY}\\ \it{
 365 Fifth Ave, New York, NY 10016, USA}\\[.5cm]
 
 \it{$^2$ The Racah Institute of Physics}\\ \it{The Hebrew University of Jerusalem} \\ \it{
Jerusalem 91904, Israel}

~\\[0.6cm]

\end{center}

\noindent 

Recent work has given a systematic way for studying the kinetics of classical weakly interacting waves beyond leading order, having analogies with renormalization in quantum field theory. An important context  is weak wave turbulence, occurring for waves which are small in magnitude and weakly interacting, such as those on the surface of the ocean. Here we continue the work of perturbatively computing correlation functions and the kinetic equation in this far-from-equilibrium state. In particular, we obtain the next-to-leading-order kinetic equation for waves with a cubic interaction.  Our main result is a simple graphical prescription for the terms in the kinetic equation, at any order in the nonlinearity.

\vfill

\end{titlepage}

\tableofcontents
~\\[-20pt]

\section{Introduction}

Kinetic equations are widely used to study equilibrium and non-equilibrium dynamics, transport, and even turbulence. The  kinetic equations  are a truncation of a vast hierarchy, involving multimode correlators. An important problem is understanding when this truncation is correct. It has long been recognized  in the context of particles (the Boltzmann equation) that corrections can be large and give qualitatively new effects \cite{PhysRev.139.A1763,PhysRevLett.25.1257, dorfman2015nonequilibrium, Dorfman}, and have been studied in a number of works \cite{green1956boltzmann, cohen1962generalization,zwanzig1963method, bogoliubov1960problems,brocas1967comparison, balescu1960irreversible, Prigogine}.  The kinetics of waves \cite{Peierls}   is particularly fascinating as it allows for the study of weak wave turbulence \cite{Falkovich}, an old subject \cite{Zakharov} undergoing a recent experimentally driven revival in a range of physical contexts, see e.g. \cite{FalconMordant} and references therein, going beyond the canonical example of ocean waves \cite{hasselmann_1962, gravity}. For waves, unlike for particles,  there hasn't been a systematic study of higher order corrections to the leading order kinetic equations.

The connection between kinetic equations and quantum field theory \cite{RS1} has  given an effective way of computing corrections, and made it evident that they are important, modifying the standard assumptions on the range of applicability of wave turbulence \cite{FR}. 
 In short, even if the coefficients of the nonlinear terms in the Hamiltonian are numerically small, intermediate states in the scattering of waves (loop diagrams) can have interacting waves with vastly different momenta. This large ratio of momenta can then overpower the smallness of the interaction coefficient, making higher order loop diagrams seemingly dominant rather than suppressed.  Summing an infinite class of loop diagrams may then give a new, renormalized, kinetic equation, which encodes phenomenon not captured by the standard leading order kinetic equation \cite{FR}. The need to compute loop diagrams, in a simple and compact way,  and potentially to high orders, is acute. This is what we seek to address, extending our previous work \cite{RS1, RS2}. Our discussion will be confined to classical kinetic theory (although we study it using tools of quantum field theory); the extension to quantum kinetic theory will be discussed in \cite{HR} and is relevant for studies of thermalization and wave turbulence in QCD, see e.g. \cite{ Arnold:2002zm, Micha:2004bv, Arnold:2005qs, Berges:2013eia, Schlichting:2019abc}.

In Sec.~\ref{sec2} we review the context of interacting waves and the leading order in nonlinearity (tree-level) correlation functions and the corresponding kinetic equation. In Sec.~\ref{sec4} we compute the one-loop correction to the correlation functions. We do this for a theory with a cubic or quartic interaction. In Sec.~\ref{sec33} we use this to give the next-to-leading order kinetic equation for a theory with a cubic interaction. This extends  the results of \cite{RS1,RS2} which studied the simplest case, of a quartic interaction. In Sec.~\ref{sec32} we give our main result: a prescription for almost immediately writing down the contribution of any Feynman diagram, at any order in the coupling, to an equal time correlation function. This turns the problem of finding higher order terms in the kinetic equation into a simple bookkeeping exercise. We conclude in Sec.~\ref{sec6}.

In Appendix~\ref{sec:Liouv} we show that an alternate method for deriving the higher order terms in the kinetic equation, by perturbatively solving the Liouville equation and performing a phase space average  \cite{RS2}, exactly reproduces -- at each order in perturbation theory -- the method used in the main body, of averaging over external Gaussian random forcing which is set to zero at the end.  In Appendix~\ref{sec:sym}  we discuss symmetries of the Hamiltonian and how it relates different terms in the kinetic equation, which is discussed further in Appendix~\ref{apc}. A few technical remarks are relegated to Appendix~\ref{technical}. 

\section{Interacting waves} \label{sec2}

Our context will be that of classical interacting waves.  This occurs in many situations, ranging from surface gravity waves to spin waves. We focus on a single field, which can, for instance, describe the height of the ocean. 
  The variables are the field and its canonical conjugate, though it is often simpler to work with a single complex field $a_p$, which is the sum of these two.
The most general Hamiltonian is a sum of $q$-body Hamiltonians,
\be \label{Hgen}
H = \sum_{q=2}^{\infty} H_q~.
\ee
In particular,  $H_2$  is the free Hamiltonian,
\be
H_2 =  \sum_p \o_p\da_p a_p~,
\ee
where $\da_p$ denotes the complex conjugate of $a_p$, $\da_p \equiv a_p^*$. 
Generally, if one is considering a theory with weak interactions, the higher order the interaction term the more suppressed it is. For this reason, it is common to focus on solely the lowest order nonvanishing term. In general, the first interaction term that dominates is the cubic term, so that the Hamiltonian is truncated to, 
\be
H = H_2 + H_3 =   \sum_p \o_p\da_p a_p + \frac{1}{2}\sum_{p_i}\( \lam_{p_1 p_2 p_3} \da_{p_1} a_{p_2} a_{p_3} + \lam^*_{p_1 p_2 p_3} a_{p_1} \da_{p_2} \da_{p_3}\)~,
\label{cubic}
\ee
where for the interaction term there is an implicit momentum conserving delta function $\delta(\v p_1 {-} \v p_2 {-} \v p_3)$. 
In sections that follow, it will be convenient to use the shorthand $\lam_{123} \equiv \lam_{p_1 p_2 p_3} $, and similarly for larger $q$.  
 In some cases the dispersion relation is such that a resonant cubic interaction is forbidden, i.e., it is not possible for the wave to have both $\v p_1 = \v p_2 + \v p_3$ and $\o_{p_1} = \o_{p_2} + \o_{p_3}$. Then the lowest interaction term is quartic. 
This is the case we focused on in \cite{RS1} and \cite{RS2} (the symmetries of the  quartic term make it slightly simpler than the cubic term),~\footnote{The cubic couplings have the symmetry $\lam_{p_1 p_2 p_3} = \lam_{p_1 p_3 p_2}$ and the quartic couplings have the symmetry $\lam_{p_1 p_2 p_3 p_4} = \lam_{p_2 p_1 p_3 p_4}  = \lam_{p_1 p_2 p_4 p_3}  = \lam_{p_3 p_4 p_1 p_2}^*$.} 
\be \label{26}
H= H_2 + H_4 =    \sum_p \o_p\da_p a_p + \sum_{p_1 \cdots p_4} \lam_{p_1 p_2 p_3 p_4} \da_{p_1}\da_{p_2} a_{p_3} a_{p_4}~.
\ee
In the main body of the text we will focus on the case of $q =3$ and $q=4$. The generalization to arbitrary $q$ is straightforward.

\subsection*{Averaging over forcing} \label{sec3}
Since this is a many-body chaotic system, one has to perform some kind of average in order to have sensible quantities. The averaging that we do here is to add Gaussian-random forcing. In particular, the equations of motion, 
\be   \label{eom1}
 \dot a_k = -i \frac{\d H}{\d \da_k}~,
 \ee
 are modified so as to add forcing, as well as  dissipation in order to absorb the flux of energy. The new equations of motion are\footnote{There are other choices of dissipation one could make, such  as $-\frac{\gamma_k}{\omega_k}\frac{\d H}{\d \da_k}$ \cite{Schubring}. In the limit of vanishing dissipation these are all equivalent.}, 
 \be   \label{eomf}
 \dot a_k = -i \frac{\d H}{\d \da_k}+ f_k(t) - \gamma_k a_k~,
 \ee
 where the forcing is drawn from a Gaussian distribution with variance $F_k$,
 \be \label{Pf2}
 P\[f\] \sim \exp\( - \int d t \sum_k \frac{|f_k(t)|^2}{F_k}\)~, \ \ \ \  \la f_k(t) f_{p}^*(t')\ra = F_k \delta(k{-}p) \delta(t{ -} t')~.
 \ee

 After computing the correlation functions,  we  take $F_k, \g_k \rightarrow 0$ with fixed $n_k$,
\be
n_k \equiv \frac{F_k}{2\g_k}~.
\ee
The reason for this notation is that $n_k$ is the occupation number of mode $k$ for the noninteracting theory. Going forward, for notational simplicity, instead of writing $F_k$, we will write $2 \g_k n_k$. 

It is important to note that we will be computing correlation functions in a stationary state. In other words, we assume that we have a stationary state and then later self-consistently pick $F_k$ and $\g_k$ to ensure this. It will turn out that there are multiple possible stationary states in the zero dissipation limit: the thermal state and the  turbulent state. However, nothing in our calculations of the correlation functions and kinetic equation is dependent on the properties of these states; we only need to assume we are computing correlators about a stationary state. 

One can compute correlation functions by solving the equations of motion perturbatively in the interaction and then averaging over forcing. 
For each term in the perturbative expansion one can associate a corresponding Wyld diagram \cite{WYLD1961143,ZakharovLvov }. 

A more streamlined method for doing this calculation was introduced in \cite{RS1}, and consists of integrating out the forcing at the outset. In particular, our classical stochastic field theory  is equivalent to a quantum field theory, with  expectation values given by a path integral for a Lagrangian that is the square of the classical, force-free, equations of motion,
\be \label{Leff}
\la \mO(a)\ra = \int  \mathcal D a \mathcal D \da\, \mO(a) \, e^{-\int dt L}~, \ \ \ \ \ L = \sum_k \frac{|E_{f=0}|^2 }{2 \g_k n_k} ~, \ \ \ \ \text{E}_{f=0} = \dot a_k + i \frac{\delta H}{\delta \da_k} +\gamma_k a_k~.
\ee

Let us break up the Hamiltonian into the free part, $H_2$, and the interacting part, $H_{int} = \sum_{q=3}^{\infty} H_q$. The Lagrangian is then, 
\be
L = \sum_k\frac{1}{2 \g_k n_k} \Big| \dot a_k +( i \o_k {+} \g_k) a_k  +i \frac{\delta H_{int}}{\delta \da_k}\Big|^2 = L_{free} + L_{int}
\ee
where 
\bea \nn
L_{free} &=& \sum_k\frac{1}{2 \g_k n_k} \Big|(\d_t + i \o_k {+}\g_k) a_k\Big|^2\\ 
L_{int} &=&\sum_k\frac{1}{2 \g_k n_k} \Big[-i(\d_t + i \o_k {+}\g_k) a_k\frac{\delta H_{int}}{\delta a_k} + \text{c.c.}\Big] + \ldots \label{Lint}~,
\eea
where  $\text{c.c.}$ denotes the complex conjugate. 
The term contained in the dots of $L_{int}$ that we have left out is the square of the absolute value of $\frac{\delta H_{int}}{\delta a_k}$; we can neglect this term provided that when evaluating Feynman diagrams we drop contact interactions (when two times collide), see Appendix \ref{technical}.
\begin{figure}[t]
\centering
\subfloat[]{
\includegraphics[width=1.3in]{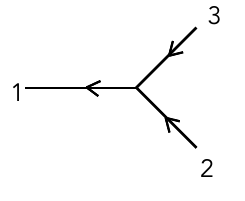}}
 \ \ \ \  \ \ \ \ \ \ \ \  \  \ \ \ \ 
\subfloat[]{
\includegraphics[width=1.2in]{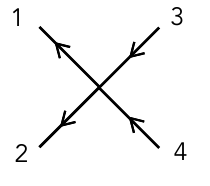}
}
\caption{ Tree-level Feynman diagrams for (a) cubic (\ref{cubicTree}) and (b) quartic interactions (\ref{quarticTree}). }  \label{tree}
\end{figure}

Let us work out the Feynman rules.  As usual, the Feynman rules are most convenient in momentum/frequency space.
The quadratic term in the Lagrangian, $L_{free}$,  gives the propagator,
\be \label{DG}
D_{k,\o}  =  n_k \frac{2\g_k}{(\o - \o_k)^2 + \g_k^2}~.
\ee
We will use shorthand $D_i \equiv D_{p_i, \o_i}$, $n_i \equiv n_{p_i}$, and $\g_i \equiv \g_{p_i}$.  Notice that in the limit of $\g_k\rightarrow 0$, the propagator becomes a delta function, 
$n_k 2\pi \delta(\o- \o_k)$. It will, however, be important to keep $\g_k$ finite until the very end of the calculation. 
It will be convenient to define,
\be \label{gi}
g_j = \frac{\o_j-{\o}_{p_j}{+}i\g_j}{2\g_{j} n_{j} i}~.
\ee
The Feynman rule for the vertex  is\footnote{Note that these Feynman rules are not symmetrized.}
\bea \nn
&&{-i\over 2}\lam_{123}\( g_1^*{-} g_{2}  {-} g_{3} \)  2\pi \delta(\o_{1;23})~, \ \ \ \ \ \  \ \ \ \  \ q=3~,\\ 
&& - i \lam_{1234} (g_1^* {+} g_2^* {-} g_3 {-} g_4) \, 2\pi \delta(\o_{12;34})~, \ \ \ \ \ q=4~, \label{vertexRule}
\eea
where we have introduced the notation $\o_{12;3} \equiv \o_1{+}\o_2{-}\o_3$ and $\o_{12;34} \equiv \o_1{+}\o_2{-}\o_3{-}\o_4$. 
The tree-level frequency space correlation functions are trivially obtained by adding external propagators and accounting for the appropriate combinatorial factors from Wick contractions, 
\bea \label{cubicTree}
\la a_{p_1, \o_1} \da_{p_2, \o_2} \da_{p_3, \o_3}\ra&=& -  i \lam_{123}(g_1^* {-} g_2 {-} g_3) D_1 D_2 D_3 \, 2\pi \delta(\o_{1;23})~, \ \ \ \ \  \ \ \  \  \  \ q=3\\
\la a_{p_1, \o_1} a_{p_2, \o_2} \da_{p_3, \o_3}\da_{p_4, \o_4}\ra&=& - 4 i \lam_{1234}(g_1^* {+} g_2 {-} g_3{-}g_4) D_1 D_2 D_3  D_4 2\pi \delta(\o_{12;34})~. \ \ q=4~, \label{quarticTree}
\eea
The corresponding Feynman diagrams are shown in Fig.~\ref{tree}. The time-space correlation functions are Fourier transforms, 
\be  \label{FT1}
\la a_{p_1}(t_1) \da_{p_2}(t_2) \da_{p_3}(t_3) \ra = \int \frac{d\o_1}{2\pi} \cdots\frac{ d \o_3}{2\pi}  e^{-i \o_1 t_1 {+}i \o_2t_2{+} i \o_3t_3 }\la a_{p_1, \o_1} \da_{p_2, \o_2} \da_{p_3, \o_3}\ra~.
\ee
In particular, the equal-time tree-level correlation functions are,
\bea \label{319}
\la a_1 \da_2 \da_3\ra &=&  \lam_{123}\frac{1}{\o_{p_2, p_3; p_1}{+}i\eps} n_1 n_2 n_3 \(\frac{1}{n_1}{-}\frac{1}{n_2}{-}\frac{1}{n_3}\)~, \\
\ \la a_1 a_2 \da_3 \da_4\ra &=&4 \lam_{1234}\frac{1}{\o_{p_3, p_4; p_1,p_2}{+}i\eps}n_1 n_2 n_3 n_4 \(\frac{1}{n_1}{+}\frac{1}{n_2}{-}\frac{1}{n_3}{-}\frac{1}{n_4}\)~,
\eea
where $\epsilon=\sum_i\gamma_i$ represents the sum over the dissipation constants appearing in the vertex. 

An important quantity is the occupation number $n_k$ of mode $k$, $n_k = \la \da_k a_k\ra$, which is governed by the kinetic equation. Through use of  the equations of motion
it can be expressed in terms of the equal-time correlation function. In particular, for $q=3$, 
\be \label{220}
\frac{\d n_k}{\d t} = \sum_{p_i} \(\delta_{k p_1} {-}\delta_{k p_2} {-} \delta_{k p_3}\) \text{Im}\( \lam_{p_1 p_2 p_3} \la \da_{p_1} a_{p_2} a_{p_3}\ra\)~,
\ee
where all operators on the left and the right are evaluated at time $t$. For the general Hamiltonian (\ref{Hgen}) this has the obvious generalization. 

Inserting the tree-level correlation function (\ref{319}) into (\ref{220}) gives the standard kinetic equation for waves \cite{Falkovich}. Our goal will be to compute higher order in the coupling corrections to this. 
In the next section we will therefore turn to computing the equal-time correlation functions perturbatively in the coupling.

\section{Loops} \label{sec4}
The tree-level correlation functions were  given in the previous section. Here we will compute their one-loop corrections, first in the case of the quartic interaction and then for the cubic interaction. 

\subsection{Quartic interaction} \label{sec41}
\begin{figure}
\centering
\subfloat[]{
\includegraphics[width=1.4in]{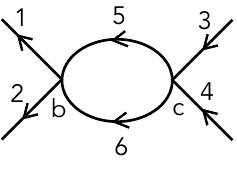}}
 \ \ \ \  \ \ \ \ \ \ \ \  \  \ \ \ \ 
\subfloat[]{
\includegraphics[width=1.4in]{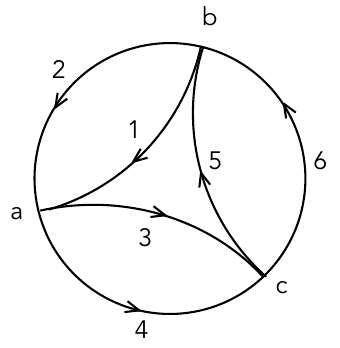}
}
\caption{ (a) A one-loop diagram contributing to the four-point function. (b) We set the external times to be equal to $t_a$, represented by the four lines meeting at the point $t_a$. }  \label{41loop}
\end{figure}

The one-loop (order $\lam^2$) correction to the tree-level four-point function for the case of $q=4$ (quartic interaction) was computed in \cite{RS1}. One of the corresponding diagrams is shown in Fig.~\ref{41loop};  there are two more diagrams with arrows in different directions which we won't discuss here.  Here we redo the calculation in a faster way, working in time space instead of frequency space, which is better suited for generalization to  higher order loops. 

Using the Feynman rules, the contribution of this diagram to the one-loop four-point function, with all external times set equal to time $t_a$, is given by,~\footnote{Our sign convention for the Fourier transform is different here than in the previous section, (\ref{FT1}), but this doesn't affect the answer.}
\be \nn
  \langle a_1 a_2 a^{\dagger}_{3}a^{\dagger}_{4}\rangle(t_a)=  8\sum_{p_5,p_6} \int\prod_{i=1}^6 \frac{d\o_i}{2\pi}D_i\, e^{i \o_{12;34}t_a} 2\pi \delta(\o_{12;56})\, 2\pi\delta(\o_{56;34})\, V+\ldots
  \ee
  \be 
 V=  -\lam_{1256}\lam_{5634} (g_1^*{+}g_2^*{-}g_5 {-}g_6)(g_5^*{+}g_6^*{-}g_3{-}g_4)~. \label{321}
   \ee
 Writing the frequency delta functions in terms of their Fourier transforms, $ 2\pi \delta(\o_{12;56})\ = \int d t_b\, e^{- i \o_{12;56}t_b}$ and $ 2\pi \delta(\o_{56;34})\ = \int d t_c\, e^{ - i \o_{56;34}t_c}$ we get, 
\be \label{323}
  \langle a_1 a_2 a^{\dagger}_{3}a^{\dagger}_{4}\rangle(t_a) =8\sum_{p_5,p_6} \int\prod_{i=1}^6 \frac{d\o_i}{2\pi}D_i\, \int d t_b d t_c\, e^{ i (\o_1 {+}\o_2) t_{ab}} e^{-i (\o_3{+}\o_4) t_{ac}}  e^{-i (\o_5 {+}\o_6)t_{cb}}\, V + \ldots~.
  \ee
  We exchange the order of integration, and first evaluate the $\o_i$ integrals. The $\o_i$ integrals are  trivial to evaluate, by closing the contour and picking up simple poles. For any particular choice of time orderings of $t_a, t_b, t_c$,  convergence imposes a unique choice of if the contours of integration should be closed in the upper or lower half-plane. In particular, the poles come from  the propagators  $D_i$ in $V$. For the integrals $\int \frac{d \o}{2\pi} e^{- i \o t_{i j}}$:  if $t_{ij}>0$ then we close in the lower-half plane, picking up the pole $\omega=\omega_{p}{-}i\gamma$.  If $t_{ij}<0$ then we close in the upper half plane, picking up the pole $\omega=\omega_{p}{{+}}i\gamma$.~\footnote{Notice that we are excluding the possibility of $t_b, t_c$ being equal; see Appendix \ref{technical}.}
  
  We split the time integrations into $3!$ regions, corresponding to the $3!$ possible time orderings of $t_a, t_b, t_c$. Unless $t_a$ is the earliest time, one finds a vanishing contribution. This leaves two possible orderings:
  
  \textbf{Region 1: $t_b>t_c>t_a$}
  
  We close the $\o_1, \o_2$ integrals in the lower half plane, and close the $\o_3, \o_4, \o_5, \o_6$ integrals in the upper half plane. The vertex factor $V$ becomes, 
  \begin{equation}
    V= \lam_{1256}\lam_{5634}\left(\frac{1}{n_1}{{+}}\frac{1}{n_2}{-}\frac{1}{n_5}{-}\frac{1}{n_6}\right)\left(\frac{1}{n_3}{{+}}\frac{1}{n_4}\right)~,
\end{equation}
and the time integral becomes, 
\be
\!\!\!\!\int^{\infty}_{t_a} \!\!\!dt_{c}\int^{\infty}_{t_c}\!\! dt_{b}\, e^{{-}i (\omega_{p_1}{+}\o_{p_2}{-2}i\gamma) t_{b a}}  e^{i(\omega_{p_3}{+}\o_{p_4}{{+}}2i\gamma) t_{ca}}
e^{i (\omega_{p_5}+\o_{p_6}{{+}}2i\gamma) t_{bc}} 
=     \frac{i}{\omega_{p_5,p_6;p_1p_2}{{+}}4i\gamma}\frac{i}{\omega_{p_3,p_4;p_1,p_2}{{+}}4i\gamma}~,\label{intc1}
\ee
so that in total we have, 
\begin{equation} \label{325}
  8\sum_{p_5} \lam_{1256}\lam_{5634} \prod_{i}^{6}n_{i}\left(\frac{1}{n_1}{{+}}\frac{1}{n_2}{-}\frac{1}{n_5}{-}\frac{1}{n_6}\right)\left(\frac{1}{n_3}{{+}}\frac{1}{n_4}\right) \frac{i}{\omega_{p_5,p_6;p_1p_2}{{+}}4i\gamma}\frac{i}{\omega_{p_3,p_4;p_1,p_2}{{+}}4i\gamma}.
\end{equation}

\paragraph{}\textbf{Region 2: $t_c>t_b>t_a$:}

We close the $\o_1,\o_2,\o_5,\o_6$ integrals in the lower half plane, and we close the $\o_3,\o_4$ integrals in the upper half plane. This vertex factor becomes
\begin{equation}
    V=-\lam_{1256}\lam_{5634}\left(\frac{1}{n_1}{{+}}\frac{1}{n_2}\right)\left(\frac{1}{n_5}{{+}}\frac{1}{n_6}{-}\frac{1}{n_3}{-}\frac{1}{n_4}\right)~,
\end{equation}
and the time integral becomes, 
\be
\!\!\!\! \int^{\infty}_{t_a}\!\! dt_{b}\int^{\infty}_{t_b}\!\! dt_{c}\, e^{{-}i (\omega_{p_1}{+}\o_{p_2}{-}2i\gamma) t_{b a}}\, e^{i(\omega_{p_3}{+}\o_{p_4}{{+}}2i\gamma) t_{ca}} 
 e^{i (\omega_{p_5}{+}\o_{p_6}{-}2i\gamma) t_{bc}} 
=
    \frac{i}{\omega_{p_3,p_4;p_5,p_6}{{+}}4i\gamma}\frac{i}{\omega_{p_3,p_4;p_1,p_2}{{+}}4i\gamma}~, \label{intc2}
\ee
so that in total we have,
\begin{equation} \label{328}
-  8\sum_{p_5} \lam_{1256}\lam_{5634} \prod_{i}^{6}n_{i}\left(\frac{1}{n_1}{{+}}\frac{1}{n_2}\right)\left(\frac{1}{n_5}{{+}}\frac{1}{n_6}{-}\frac{1}{n_3}{-}\frac{1}{n_4}\right)  \frac{i}{\omega_{p_3,p_4;p_5,p_6}{{+}}4i\gamma}\frac{i}{\omega_{p_3,p_4;p_1,p_2}{{+}}4i\gamma}~.
\end{equation}

The sum of (\ref{325}) and (\ref{328}) reproduces what we found earlier in \cite{RS1} (see Eq.~4.25 and use the identity A.8 in \cite{RS2}).~\footnote{There is an additional Feynman diagram that contributes to the one-loop kinetic equation, whose contribution simply involves transforming the diagram we discussed by $2\leftrightarrow -3$ and $6\rightarrow -6$, see \cite{RS1}.}

\subsection{Cubic interaction} \label{cubicMain}

Consider now the equal-time three-point function $\la a_1 \da_2 \da_3\ra(t_a)$
 arising from a cubic interaction. The tree-level correlation function, at order $\lam$, was given earlier in (\ref{319}). At order $\lam^2$ all Feynman diagrams vanish. At order $\lam^3$ one has two kinds of diagrams: the ``tetrahedron'' diagram, which can be viewed as a renormalization of the cubic interaction vertex,  and a loop diagram arising from renormalization of the propagator. 
 
 \subsubsection*{Vertex renormalization: tetrahedron diagram}
 \begin{figure}[h]
\centering
\subfloat[]{

\tikzset{every picture/.style={line width=0.75pt}} 

\begin{tikzpicture}[x=0.75pt,y=0.75pt,yscale=-1,xscale=1]

\draw    (127.6,116.88) -- (81.73,116.88) ;
\draw [shift={(98.67,116.88)}, rotate = 360] [color={rgb, 255:red, 0; green, 0; blue, 0 }  ][line width=0.75]    (10.93,-3.29) .. controls (6.95,-1.4) and (3.31,-0.3) .. (0,0) .. controls (3.31,0.3) and (6.95,1.4) .. (10.93,3.29)   ;
\draw    (177.6,78.47) -- (127.6,116.88) ;
\draw [shift={(147.84,101.33)}, rotate = 322.47] [color={rgb, 255:red, 0; green, 0; blue, 0 }  ][line width=0.75]    (10.93,-3.29) .. controls (6.95,-1.4) and (3.31,-0.3) .. (0,0) .. controls (3.31,0.3) and (6.95,1.4) .. (10.93,3.29)   ;
\draw    (176.27,152.47) -- (127.6,116.88) ;
\draw [shift={(147.09,131.13)}, rotate = 36.18] [color={rgb, 255:red, 0; green, 0; blue, 0 }  ][line width=0.75]    (10.93,-3.29) .. controls (6.95,-1.4) and (3.31,-0.3) .. (0,0) .. controls (3.31,0.3) and (6.95,1.4) .. (10.93,3.29)   ;
\draw    (176.27,78.47) -- (176.27,152.47) ;
\draw [shift={(176.27,121.47)}, rotate = 270] [color={rgb, 255:red, 0; green, 0; blue, 0 }  ][line width=0.75]    (10.93,-3.29) .. controls (6.95,-1.4) and (3.31,-0.3) .. (0,0) .. controls (3.31,0.3) and (6.95,1.4) .. (10.93,3.29)   ;
\draw    (207.6,55.13) -- (177.6,78.47) ;
\draw [shift={(187.86,70.48)}, rotate = 322.13] [color={rgb, 255:red, 0; green, 0; blue, 0 }  ][line width=0.75]    (10.93,-3.29) .. controls (6.95,-1.4) and (3.31,-0.3) .. (0,0) .. controls (3.31,0.3) and (6.95,1.4) .. (10.93,3.29)   ;
\draw    (203.6,173.13) -- (176.27,152.47) ;
\draw [shift={(185.15,159.18)}, rotate = 37.09] [color={rgb, 255:red, 0; green, 0; blue, 0 }  ][line width=0.75]    (10.93,-3.29) .. controls (6.95,-1.4) and (3.31,-0.3) .. (0,0) .. controls (3.31,0.3) and (6.95,1.4) .. (10.93,3.29)   ;

\draw (98,125) node [anchor=north west][inner sep=0.75pt]   [align=left] {1};
\draw (210,45.67) node [anchor=north west][inner sep=0.75pt]   [align=left] {3};
\draw (210,167.67) node [anchor=north west][inner sep=0.75pt]   [align=left] {2};
\draw (144,147) node [anchor=north west][inner sep=0.75pt]   [align=left] {4};
\draw (138.67,77.67) node [anchor=north west][inner sep=0.75pt]   [align=left] {5};
\draw (188.67,107) node [anchor=north west][inner sep=0.75pt]   [align=left] {6};
\draw (120.67,121.67) node [anchor=north west][inner sep=0.75pt]   [align=left] {c};
\draw (169.33,61.67) node [anchor=north west][inner sep=0.75pt]   [align=left] {b};
\draw (169.33,160.33) node [anchor=north west][inner sep=0.75pt]   [align=left] {d};

\end{tikzpicture}}
 \ \ \ \  \ \ \ \ \ \ \ \  \  \ \ \ \ 
\subfloat[]{\includegraphics[width=2in]{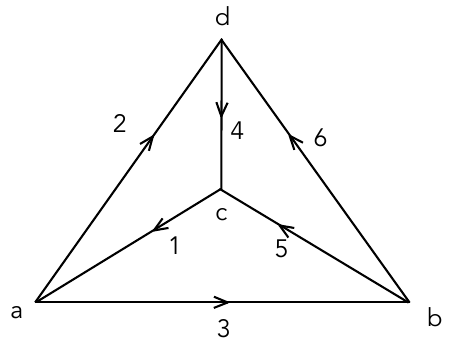}
}
\caption{ (a) A one-loop diagram contributing to the three-point function. (b) We set the external times to be equal to $t_a$, represented by the three lines meeting at the point $t_a$. We refer to this as the tetrahedron diagram }  \label{31loop}
\end{figure}
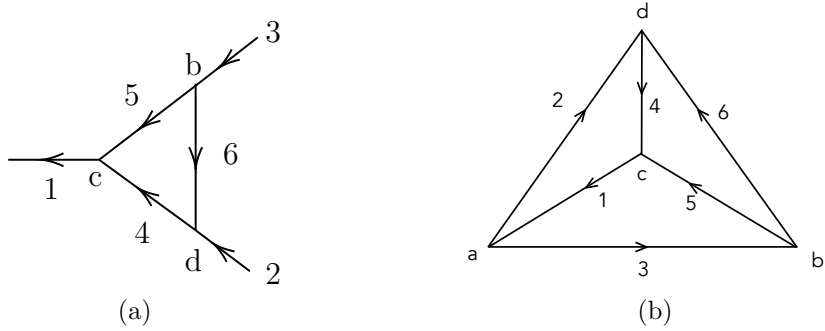

 We start with tetrahedron  diagram shown in Fig.~\ref{31loop}. 
Evaluating the diagram in an analogous way to the diagram in the previous section, we find  its contribution to the equal-time three-point function $\la a_1 \da_2 \da_3\ra$ is equal to,
 \bml  \label{329}
2\sum_{p_4,p_5,p_6} \lam_{426} \lam_{356}^* \lam_{145} \prod_{i=1}^6 n_i  \frac{1}{\o_{p_2,p_3;p_1}{+}i\epsilon}\\
\Big\{ \frac{1}{n_1} \frac{1}{\o_{p_2,p_3;p_4, p_5}{+}i\eps}\Big[ \frac{1}{\o_{p_3;p_5,p_6}{+}i\eps} \Big(\frac{1}{n_4}{-}\frac{1}{n_2}\Big)\Big(\frac{1}{n_5} {+}\frac{1}{n_6}{-}\frac{1}{n_3}\Big) + \frac{1}{\o_{p_2,p_6;p_4}{+}i\eps} \Big(\frac{1}{n_5}{-}\frac{1}{n_3}\Big)\Big(\frac{1}{n_4} {-}\frac{1}{n_2}{-}\frac{1}{n_6}\Big) \Big] \\
 -\frac{1}{n_2} \frac{1}{\o_{p_3,p_4;p_1,p_6}{+}i\eps}\Big[ \frac{1}{\o_{p_3;p_5,p_6}{+}i\eps} \Big(\frac{1}{n_1}{-}\frac{1}{n_4}\Big)\Big(\frac{1}{n_5} {+}\frac{1}{n_6}{-}\frac{1}{n_3}\Big) + \frac{1}{\o_{p_4,p_5;p_1}{+}i\eps} \Big(\frac{1}{n_6}{-}\frac{1}{n_3}\Big)\Big(\frac{1}{n_1} {-}\frac{1}{n_4}{-}\frac{1}{n_5}\Big) \Big] \\
 -\frac{1}{n_3} \frac{1}{\o_{p_2,p_5,p_6;p_1}{+}i\eps}\Big[ \frac{1}{\o_{p_4,p_5;p_1}{+}i\eps} \Big(\frac{-1}{n_2}{-}\frac{1}{n_6}\Big)\Big(\frac{1}{n_1} {-}\frac{1}{n_4}{-}\frac{1}{n_5}\Big) + \frac{1}{\o_{p_2,p_6;p_4}{+}i\eps} \Big(\frac{1}{n_1}{-}\frac{1}{n_5}\Big)\Big(\frac{-1}{n_2} {+}\frac{1}{n_4}{-}\frac{1}{n_6}\Big) \Big] \Big\}~,
\end{multline}
where the six terms arise from the $3!$ possible orderings of the times  $t_b, t_c, t_d > t_a$. In particular, the time orderings of the six terms are: $t_b{>}t_d{>}t_c$, $t_d{>}t_b{>}t_c$,  $t_b{>}t_c{>}t_d$, $t_c{>}t_b{>}t_d$, $t_c{>} t_d{>}t_b$, $t_d{>}t_c{>}t_b$,  from first to last, respectively. In fact, as a result of the symmetries of the tetrahedron, each of the six terms can be obtained by a symmetry transformation of one another, see Appendix~\ref{sec:sym}.

 In addition to this diagram, there is the same tetrahedron diagram but with the arrow on line $5$ reversed. The result for the latter can be obtained from the former in a simple way, from symmetry, and is also discussed in Appendix~\ref{sec:sym}.

\subsubsection*{Propagator renormalization }  \label{sec5}

 \begin{figure}[h]
\centering
\subfloat[]{

\begin{tikzpicture}[x=0.75pt,y=0.75pt,yscale=-1,xscale=1]

\draw    (136.76,57.54) -- (95.4,57.54) ;
\draw [shift={(110.08,57.54)}, rotate = 360] [color={rgb, 255:red, 0; green, 0; blue, 0 }  ][line width=0.75]    (10.93,-3.29) .. controls (6.95,-1.4) and (3.31,-0.3) .. (0,0) .. controls (3.31,0.3) and (6.95,1.4) .. (10.93,3.29)   ;
\draw   (136.76,55.44) .. controls (136.76,40.65) and (148.09,28.65) .. (162.08,28.65) .. controls (176.06,28.65) and (187.4,40.65) .. (187.4,55.44) .. controls (187.4,70.23) and (176.06,82.23) .. (162.08,82.23) .. controls (148.09,82.23) and (136.76,70.23) .. (136.76,55.44) -- cycle ;
\draw    (159.73,82.23) -- (167.33,77.67) ;
\draw    (159.73,82.23) -- (167.13,86.48) ;
\draw    (221.05,57.93) -- (187.71,57.93) ;
\draw [shift={(198.38,57.93)}, rotate = 360] [color={rgb, 255:red, 0; green, 0; blue, 0 }  ][line width=0.75]    (10.93,-3.29) .. controls (6.95,-1.4) and (3.31,-0.3) .. (0,0) .. controls (3.31,0.3) and (6.95,1.4) .. (10.93,3.29)   ;
\draw    (259,33.2) -- (221.05,57.93) ;
\draw [shift={(235,48.84)}, rotate = 326.91] [color={rgb, 255:red, 0; green, 0; blue, 0 }  ][line width=0.75]    (10.93,-3.29) .. controls (6.95,-1.4) and (3.31,-0.3) .. (0,0) .. controls (3.31,0.3) and (6.95,1.4) .. (10.93,3.29)   ;
\draw    (258,82.2) -- (221.05,57.93) ;
\draw [shift={(234.51,66.77)}, rotate = 33.3] [color={rgb, 255:red, 0; green, 0; blue, 0 }  ][line width=0.75]    (10.93,-3.29) .. controls (6.95,-1.4) and (3.31,-0.3) .. (0,0) .. controls (3.31,0.3) and (6.95,1.4) .. (10.93,3.29)   ;
\draw    (158.73,29.23) -- (168,25.33) ;
\draw    (158.73,29.23) -- (168,33.33) ;

\draw (79,59) node [anchor=north west][inner sep=0.75pt]   [align=left] {1};
\draw (262,76) node [anchor=north west][inner sep=0.75pt]   [align=left] {2};
\draw (264,23) node [anchor=north west][inner sep=0.75pt]   [align=left] {3};
\draw (203,65) node [anchor=north west][inner sep=0.75pt]   [align=left] {4};
\draw (159,5) node [anchor=north west][inner sep=0.75pt]   [align=left] {5};
\draw (158,93) node [anchor=north west][inner sep=0.75pt]   [align=left] {6};
\draw (125,62) node [anchor=north west][inner sep=0.75pt]   [align=left] {d};
\draw (189.4,58.44) node [anchor=north west][inner sep=0.75pt]   [align=left] {c};
\draw (217,59) node [anchor=north west][inner sep=0.75pt]   [align=left] {b};
\end{tikzpicture}
}
 \ \ \ \  \ \ \ \ \ \ \ \  \  \ \ \ \ 
\subfloat[]{\includegraphics[width=2in]{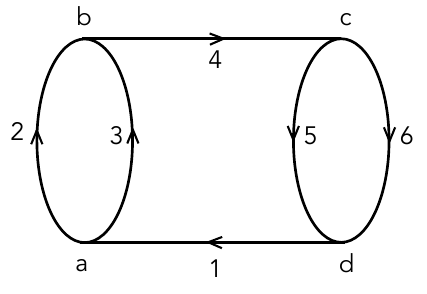}
}
\caption{  (a) A one-loop diagram contributing to the three-point function. (b) We set the external times to be equal to $t_a$, represented by the three lines meeting at the point $t_a$.  }  \label{Threepillow}
\end{figure}
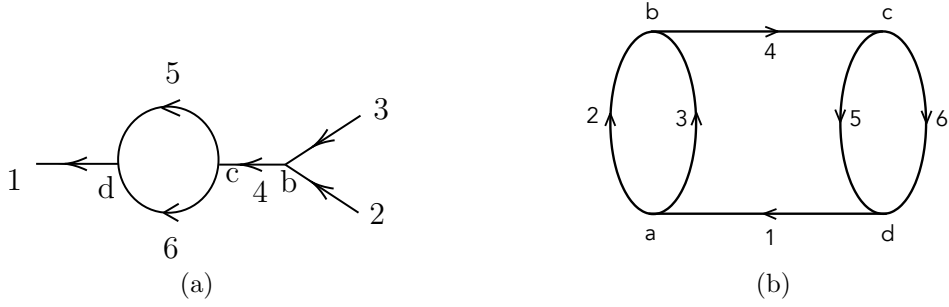

\begin{figure}
\centering

\end{figure}

Now we look at the loop diagram shown in Fig.~\ref{Threepillow} which can be viewed as arising from propagator renormalization. We find that its contribution to the equal-time three-point function is,
\bml \label{threepillowa}
\frac{-1}{2}\sum_{p_4,p_5,p_6} \lam_{423} \lam^*_{456}\lam_{156} \prod_{i=1}^6 n_i\, \frac{1}{\o_{p_2,p_3; p_1}{+}i\eps}\Big\{ \frac{1}{\o_{p_2,p_3; p_4}{+}i\eps} \frac{1}{\o_{p_2, p_3; p_5, p_6}{+}i\eps}  \frac{1}{n_1}\(\frac{1}{n_5}{+}\frac{1}{n_6}\)\(\frac{1}{n_2}{+}\frac{1}{n_3}{-}\frac{1}{n_4}\)\\
- \frac{1}{\o_{p_4;p_5,p_6}{+}i\eps} \(  \frac{1}{\o_{p_2, p_3;p_5,p_6}{+}i\eps}+ \frac{1}{\o_{p_4;p_1}{+}i\eps}\)\frac{1}{n_1} \(\frac{1}{n_2}{+}\frac{1}{n_3}\)  \(\frac{1}{n_4}{-}\frac{1}{n_5}{-}\frac{1}{n_6}\)\\
+\frac{1}{\o_{p_5, p_6;p_1}{+}i\eps}\frac{1}{\o_{p_4;p_1}{+}i\eps} \frac{1}{n_4}   \(\frac{1}{n_2}{+}\frac{1}{n_3}\)  \(\frac{1}{n_5}{+}\frac{1}{n_6}{-}\frac{1}{n_1}\)
 \Big\}~,
\end{multline}
where the terms correspond to the time orderings $t_b{>}t_c{>}t_d$, $t_c{>}t_b{>}t_d$, $t_c{>}t_d{>}t_b$, $t_d{>}t_c{>}t_b$ from  first to last, respectively.~\footnote{The combinatorial factor of $1/2$ out front is: $1/8$ because each vertex comes with a $1/2$, then a combinatorial factor of $2$ from the loop integral, and a factor of $2$ from exchanging $2$ and $3$ when contracting with the external legs.} 

There are several more diagrams of this kind, which can be viewed as a symmetry transformation of this diagram, as we will see below. In fact, in (\ref{threepillowa}) one should, by momentum conservation, set $p_4=p_1$. This gives some terms that are divergent. Specifically, (\ref{threepillowa}) gives the term, 
\be \label{pillowdiv1}
\frac{-1}{i\eps}\frac{1}{2}\sum_{1, \ldots, 6} \lam_{123}|\lam_{156}|^2 \prod_{i=1}^6 n_i\, \frac{1}{\o_{p_2,p_3; p_1}{+}i\eps} \frac{1}{n_1} \(\frac{1}{n_2}{+}\frac{1}{n_3}\) 
2\pi i \delta(\o_{p_1;p_5,p_6})\(\frac{1}{n_1}{-}\frac{1}{n_5}{-}\frac{1}{n_6}\)~.
\ee 
However, these terms cancel when including the diagram Fig.~\ref{PropRenorm}(b) that will be shown in the next section, provided that one uses that the state is stationary, as governed by the leading order kinetic equation. In particular, adding to (\ref{pillowdiv1}) the corresponding divergent contribution of the diagram Fig.~\ref{PropRenorm}(b) gives, 
\bml \label{pillowdiv2}
\frac{-1}{i\eps}\frac{1}{2}\sum_{1, \ldots, 6} \lam_{123}\prod_{i=1}^6 n_i\, \frac{1}{\o_{p_2,p_3; p_1}{+}i\eps} \frac{1}{n_1} \(\frac{1}{n_2}{+}\frac{1}{n_3}\) \\
\[
 |\lam_{156}|^2 2\pi i \delta(\o_{p_1;p_5,p_6})\(\frac{1}{n_1}{-}\frac{1}{n_5}{-}\frac{1}{n_6}\) +2 |\lam_{615}|^2 2\pi i \delta(\o_{p_1 p_5;p_6})\(\frac{1}{n_1}{+}\frac{1}{n_5}{-}\frac{1}{n_6}\) \]~.
\end{multline}
However, the term in brackets vanishes by the tree-level kinetic equation.  This is a useful illustration of the fact that all our computations are of equal-time correlation functions, assuming a stationary state. 

\subsection{Next-to-leading order kinetic equation for cubic interactions} \label{sec33}
We are now ready to write down the kinetic equation to next-to-leading order. It follows from the equal-time three-point function, via (\ref{220}). The complex conjugate version of this equation is,
\be
\frac{\d n_k}{\d t} =- \sum_{p_1, p_2, p_3} \(\delta_{k p_1} {-}\delta_{k p_2} {-} \delta_{k p_3}\) \text{Im}\( \lam^*_{123} \la a_{p_1} \da_{p_2} \da_{p_3}\ra\)~.
\ee
The three-point function consists of an infinite sequence of terms, grouped by their order in $\lambda$, and we go up to order $\lam^3$, 
\bea
\la a_{p_1} \da_{p_2} \da_{p_3}\ra &=& \la a_{p_1} \da_{p_2} \da_{p_3}\ra_{\lam} + \la a_{p_1} \da_{p_2} \da_{p_3}\ra_{\lam^3}\\
  \la a_{p_1} \da_{p_2} \da_{p_3}\ra_{\lam} &=& \lam_{123}\frac{1}{\o_{p_2, p_3; p_1}{+}i\eps} n_1 n_2 n_3 \(\frac{1}{n_1}{-}\frac{1}{n_2}{-}\frac{1}{n_3}\) \\
\la a_{p_1} \da_{p_2} \da_{p_3}\ra_{\lam^3} &=& V_a + V_b + P_a + P_b +P_c + P_d +P_e
\eea

 \begin{figure}[h]
\centering
\subfloat[]{

\begin{tikzpicture}[x=0.75pt,y=0.75pt,yscale=-1,xscale=1]

\draw    (136.76,57.54) -- (95.4,57.54) ;
\draw [shift={(110.08,57.54)}, rotate = 360] [color={rgb, 255:red, 0; green, 0; blue, 0 }  ][line width=0.75]    (10.93,-3.29) .. controls (6.95,-1.4) and (3.31,-0.3) .. (0,0) .. controls (3.31,0.3) and (6.95,1.4) .. (10.93,3.29)   ;
\draw   (136.76,55.44) .. controls (136.76,40.65) and (148.09,28.65) .. (162.08,28.65) .. controls (176.06,28.65) and (187.4,40.65) .. (187.4,55.44) .. controls (187.4,70.23) and (176.06,82.23) .. (162.08,82.23) .. controls (148.09,82.23) and (136.76,70.23) .. (136.76,55.44) -- cycle ;
\draw    (159.73,82.23) -- (167.33,77.67) ;
\draw    (159.73,82.23) -- (167.13,86.48) ;
\draw    (221.05,57.93) -- (187.71,57.93) ;
\draw [shift={(198.38,57.93)}, rotate = 360] [color={rgb, 255:red, 0; green, 0; blue, 0 }  ][line width=0.75]    (10.93,-3.29) .. controls (6.95,-1.4) and (3.31,-0.3) .. (0,0) .. controls (3.31,0.3) and (6.95,1.4) .. (10.93,3.29)   ;
\draw    (259,33.2) -- (221.05,57.93) ;
\draw [shift={(235,48.84)}, rotate = 326.91] [color={rgb, 255:red, 0; green, 0; blue, 0 }  ][line width=0.75]    (10.93,-3.29) .. controls (6.95,-1.4) and (3.31,-0.3) .. (0,0) .. controls (3.31,0.3) and (6.95,1.4) .. (10.93,3.29)   ;
\draw    (258,82.2) -- (221.05,57.93) ;
\draw [shift={(234.51,66.77)}, rotate = 33.3] [color={rgb, 255:red, 0; green, 0; blue, 0 }  ][line width=0.75]    (10.93,-3.29) .. controls (6.95,-1.4) and (3.31,-0.3) .. (0,0) .. controls (3.31,0.3) and (6.95,1.4) .. (10.93,3.29)   ;
\draw    (158.73,29.23) -- (168,25.33) ;
\draw    (158.73,29.23) -- (168,33.33) ;

\draw (79,59) node [anchor=north west][inner sep=0.75pt]   [align=left] {1};
\draw (262,76) node [anchor=north west][inner sep=0.75pt]   [align=left] {2};
\draw (264,23) node [anchor=north west][inner sep=0.75pt]   [align=left] {3};
\draw (203,65) node [anchor=north west][inner sep=0.75pt]   [align=left] {4};
\draw (159,5) node [anchor=north west][inner sep=0.75pt]   [align=left] {5};
\draw (158,93) node [anchor=north west][inner sep=0.75pt]   [align=left] {6};

\end{tikzpicture}
}
 \ \ \ \  
\subfloat[]{

\begin{tikzpicture}[x=0.75pt,y=0.75pt,yscale=-1,xscale=1]

\draw    (136.76,57.54) -- (95.4,57.54) ;
\draw [shift={(110.08,57.54)}, rotate = 360] [color={rgb, 255:red, 0; green, 0; blue, 0 }  ][line width=0.75]    (10.93,-3.29) .. controls (6.95,-1.4) and (3.31,-0.3) .. (0,0) .. controls (3.31,0.3) and (6.95,1.4) .. (10.93,3.29)   ;
\draw   (136.76,55.44) .. controls (136.76,40.65) and (148.09,28.65) .. (162.08,28.65) .. controls (176.06,28.65) and (187.4,40.65) .. (187.4,55.44) .. controls (187.4,70.23) and (176.06,82.23) .. (162.08,82.23) .. controls (148.09,82.23) and (136.76,70.23) .. (136.76,55.44) -- cycle ;
\draw    (159.73,82.23) -- (167.33,77.67) ;
\draw    (159.73,82.23) -- (167.13,86.48) ;
\draw    (221.05,57.93) -- (187.71,57.93) ;
\draw [shift={(198.38,57.93)}, rotate = 360] [color={rgb, 255:red, 0; green, 0; blue, 0 }  ][line width=0.75]    (10.93,-3.29) .. controls (6.95,-1.4) and (3.31,-0.3) .. (0,0) .. controls (3.31,0.3) and (6.95,1.4) .. (10.93,3.29)   ;
\draw    (259,33.2) -- (221.05,57.93) ;
\draw [shift={(235,48.84)}, rotate = 326.91] [color={rgb, 255:red, 0; green, 0; blue, 0 }  ][line width=0.75]    (10.93,-3.29) .. controls (6.95,-1.4) and (3.31,-0.3) .. (0,0) .. controls (3.31,0.3) and (6.95,1.4) .. (10.93,3.29)   ;
\draw    (258,82.2) -- (221.05,57.93) ;
\draw [shift={(234.51,66.77)}, rotate = 33.3] [color={rgb, 255:red, 0; green, 0; blue, 0 }  ][line width=0.75]    (10.93,-3.29) .. controls (6.95,-1.4) and (3.31,-0.3) .. (0,0) .. controls (3.31,0.3) and (6.95,1.4) .. (10.93,3.29)   ;
\draw    (168,29.23) -- (158.73,25.33) ;
\draw    (168,29.23) -- (158.73,33.33) ;

\draw (79,59) node [anchor=north west][inner sep=0.75pt]   [align=left] {1};
\draw (262,76) node [anchor=north west][inner sep=0.75pt]   [align=left] {2};
\draw (264,23) node [anchor=north west][inner sep=0.75pt]   [align=left] {3};
\draw (203,65) node [anchor=north west][inner sep=0.75pt]   [align=left] {4};
\draw (159,5) node [anchor=north west][inner sep=0.75pt]   [align=left] {5};
\draw (158,93) node [anchor=north west][inner sep=0.75pt]   [align=left] {6};

\end{tikzpicture}
}
 \ \ \ 
\subfloat[]{

\begin{tikzpicture}[x=0.75pt,y=0.75pt,yscale=-1,xscale=1]

\draw    (136.76,56.44) -- (92,56.44) ;
\draw [shift={(121.38,56.44)}, rotate = 180] [color={rgb, 255:red, 0; green, 0; blue, 0 }  ][line width=0.75]    (10.93,-3.29) .. controls (6.95,-1.4) and (3.31,-0.3) .. (0,0) .. controls (3.31,0.3) and (6.95,1.4) .. (10.93,3.29)   ;
\draw   (136.76,55.44) .. controls (136.76,40.65) and (148.09,28.65) .. (162.08,28.65) .. controls (176.06,28.65) and (187.4,40.65) .. (187.4,55.44) .. controls (187.4,70.23) and (176.06,82.23) .. (162.08,82.23) .. controls (148.09,82.23) and (136.76,70.23) .. (136.76,55.44) -- cycle ;
\draw    (167.33,82.23) -- (159.73,77.67) ;
\draw    (167.33,82.23) -- (159.93,86.48) ;

\draw    (187.4,57.93) -- (221.05,57.93) ;
\draw [shift={(210.23,57.93)}, rotate = 180] [color={rgb, 255:red, 0; green, 0; blue, 0 }  ][line width=0.75]    (10.93,-3.29) .. controls (6.95,-1.4) and (3.31,-0.3) .. (0,0) .. controls (3.31,0.3) and (6.95,1.4) .. (10.93,3.29)   ;
\draw    (221.05,57.93) -- (261,32) ;
\draw [shift={(246.06,41.7)}, rotate = 147.02] [color={rgb, 255:red, 0; green, 0; blue, 0 }  ][line width=0.75]    (10.93,-3.29) .. controls (6.95,-1.4) and (3.31,-0.3) .. (0,0) .. controls (3.31,0.3) and (6.95,1.4) .. (10.93,3.29)   ;
\draw    (258,82.2) -- (221.05,57.93) ;
\draw [shift={(234.51,66.77)}, rotate = 33.3] [color={rgb, 255:red, 0; green, 0; blue, 0 }  ][line width=0.75]    (10.93,-3.29) .. controls (6.95,-1.4) and (3.31,-0.3) .. (0,0) .. controls (3.31,0.3) and (6.95,1.4) .. (10.93,3.29)   ;
\draw    (158.73,29.23) -- (168,25.33) ;
\draw    (158.73,29.23) -- (168,33.33) ;

\draw (79,59) node [anchor=north west][inner sep=0.75pt]   [align=left] {3};
\draw (262,76) node [anchor=north west][inner sep=0.75pt]   [align=left] {2};
\draw (264,23) node [anchor=north west][inner sep=0.75pt]   [align=left] {1};
\draw (203,65) node [anchor=north west][inner sep=0.75pt]   [align=left] {4};
\draw (159,5) node [anchor=north west][inner sep=0.75pt]   [align=left] {5};
\draw (158,93) node [anchor=north west][inner sep=0.75pt]   [align=left] {6};

\end{tikzpicture}
} \ \ \ \  \ \ \ \ \ \ \ \  \  \ \ \ \ 
\subfloat[]{

\begin{tikzpicture}[x=0.75pt,y=0.75pt,yscale=-1,xscale=1]

\draw    (136.76,56.44) -- (92,56.44) ;
\draw [shift={(121.38,56.44)}, rotate = 180] [color={rgb, 255:red, 0; green, 0; blue, 0 }  ][line width=0.75]    (10.93,-3.29) .. controls (6.95,-1.4) and (3.31,-0.3) .. (0,0) .. controls (3.31,0.3) and (6.95,1.4) .. (10.93,3.29)   ;
\draw   (136.76,55.44) .. controls (136.76,40.65) and (148.09,28.65) .. (162.08,28.65) .. controls (176.06,28.65) and (187.4,40.65) .. (187.4,55.44) .. controls (187.4,70.23) and (176.06,82.23) .. (162.08,82.23) .. controls (148.09,82.23) and (136.76,70.23) .. (136.76,55.44) -- cycle ;
\draw    (167.33,82.23) -- (159.73,77.67) ;
\draw    (167.33,82.23) -- (159.93,86.48) ;

\draw    (187.4,57.93) -- (221.05,57.93) ;
\draw [shift={(210.23,57.93)}, rotate = 180] [color={rgb, 255:red, 0; green, 0; blue, 0 }  ][line width=0.75]    (10.93,-3.29) .. controls (6.95,-1.4) and (3.31,-0.3) .. (0,0) .. controls (3.31,0.3) and (6.95,1.4) .. (10.93,3.29)   ;
\draw    (221.05,57.93) -- (261,32) ;
\draw [shift={(246.06,41.7)}, rotate = 147.02] [color={rgb, 255:red, 0; green, 0; blue, 0 }  ][line width=0.75]    (10.93,-3.29) .. controls (6.95,-1.4) and (3.31,-0.3) .. (0,0) .. controls (3.31,0.3) and (6.95,1.4) .. (10.93,3.29)   ;
\draw    (258,82.2) -- (221.05,57.93) ;
\draw [shift={(234.51,66.77)}, rotate = 33.3] [color={rgb, 255:red, 0; green, 0; blue, 0 }  ][line width=0.75]    (10.93,-3.29) .. controls (6.95,-1.4) and (3.31,-0.3) .. (0,0) .. controls (3.31,0.3) and (6.95,1.4) .. (10.93,3.29)   ;
\draw    (168,29.23) -- (158.73,25.33) ;
\draw    (168,29.23) -- (158.73,33.33) ;

\draw (79,59) node [anchor=north west][inner sep=0.75pt]   [align=left] {3};
\draw (262,76) node [anchor=north west][inner sep=0.75pt]   [align=left] {2};
\draw (264,23) node [anchor=north west][inner sep=0.75pt]   [align=left] {1};
\draw (203,65) node [anchor=north west][inner sep=0.75pt]   [align=left] {4};
\draw (159,5) node [anchor=north west][inner sep=0.75pt]   [align=left] {5};
\draw (158,93) node [anchor=north west][inner sep=0.75pt]   [align=left] {6};

\end{tikzpicture}
}\ \ \ \ \
\subfloat[]{
\begin{tikzpicture}[x=0.75pt,y=0.75pt,yscale=-1,xscale=1]

\draw    (136.76,56.44) -- (92,56.44) ;
\draw [shift={(121.38,56.44)}, rotate = 180] [color={rgb, 255:red, 0; green, 0; blue, 0 }  ][line width=0.75]    (10.93,-3.29) .. controls (6.95,-1.4) and (3.31,-0.3) .. (0,0) .. controls (3.31,0.3) and (6.95,1.4) .. (10.93,3.29)   ;
\draw   (136.76,55.44) .. controls (136.76,40.65) and (148.09,28.65) .. (162.08,28.65) .. controls (176.06,28.65) and (187.4,40.65) .. (187.4,55.44) .. controls (187.4,70.23) and (176.06,82.23) .. (162.08,82.23) .. controls (148.09,82.23) and (136.76,70.23) .. (136.76,55.44) -- cycle ;
\draw    (167.33,82.23) -- (159.73,77.67) ;
\draw    (167.33,82.23) -- (159.93,86.48) ;

\draw    (221.05,57.93) -- (187.4,57.93) ;
\draw [shift={(198.23,57.93)}, rotate = 360] [color={rgb, 255:red, 0; green, 0; blue, 0 }  ][line width=0.75]    (10.93,-3.29) .. controls (6.95,-1.4) and (3.31,-0.3) .. (0,0) .. controls (3.31,0.3) and (6.95,1.4) .. (10.93,3.29)   ;
\draw    (221.05,57.93) -- (261,32) ;
\draw [shift={(246.06,41.7)}, rotate = 147.02] [color={rgb, 255:red, 0; green, 0; blue, 0 }  ][line width=0.75]    (10.93,-3.29) .. controls (6.95,-1.4) and (3.31,-0.3) .. (0,0) .. controls (3.31,0.3) and (6.95,1.4) .. (10.93,3.29)   ;
\draw    (258,82.2) -- (221.05,57.93) ;
\draw [shift={(234.51,66.77)}, rotate = 33.3] [color={rgb, 255:red, 0; green, 0; blue, 0 }  ][line width=0.75]    (10.93,-3.29) .. controls (6.95,-1.4) and (3.31,-0.3) .. (0,0) .. controls (3.31,0.3) and (6.95,1.4) .. (10.93,3.29)   ;
\draw    (158.73,29.23) -- (168,25.33) ;
\draw    (158.73,29.23) -- (168,33.33) ;

\draw (79,59) node [anchor=north west][inner sep=0.75pt]   [align=left] {3};
\draw (262,76) node [anchor=north west][inner sep=0.75pt]   [align=left] {2};
\draw (264,23) node [anchor=north west][inner sep=0.75pt]   [align=left] {1};
\draw (203,65) node [anchor=north west][inner sep=0.75pt]   [align=left] {4};
\draw (159,5) node [anchor=north west][inner sep=0.75pt]   [align=left] {5};
\draw (158,93) node [anchor=north west][inner sep=0.75pt]   [align=left] {6};
\end{tikzpicture}}
\caption{ The different propagator renormalization diagrams contributing to the three-point functions. All the diagrams can be obtained from $a$ (whose contribution is denoted by $P_a$) through symmetry transformations, see Appendix~\ref{sec:sym}. In particular, we get  $P_b$ from $P_a$ by sending $5\rightarrow -5$. We get $P_c$ and $P_d$ from $P_a$ and $P_b$, respectively, by sending $1\leftrightarrow -3$ and $6\rightarrow-6$ (and also $4\rightarrow -4$, however $4$ in $a$ is just $1$). Finally, we get diagram $e$ from diagram $a$ by sending $1\leftrightarrow -3$ and $6\rightarrow -6$.  Note that diagrams $c,d,e$ have a propagator renormalization for line $3$. We could have alternatively put it on line $2$; we account for this by adding a factor of $2$ for $P_c, P_d, P_3$. }  \label{PropRenorm}
\end{figure}
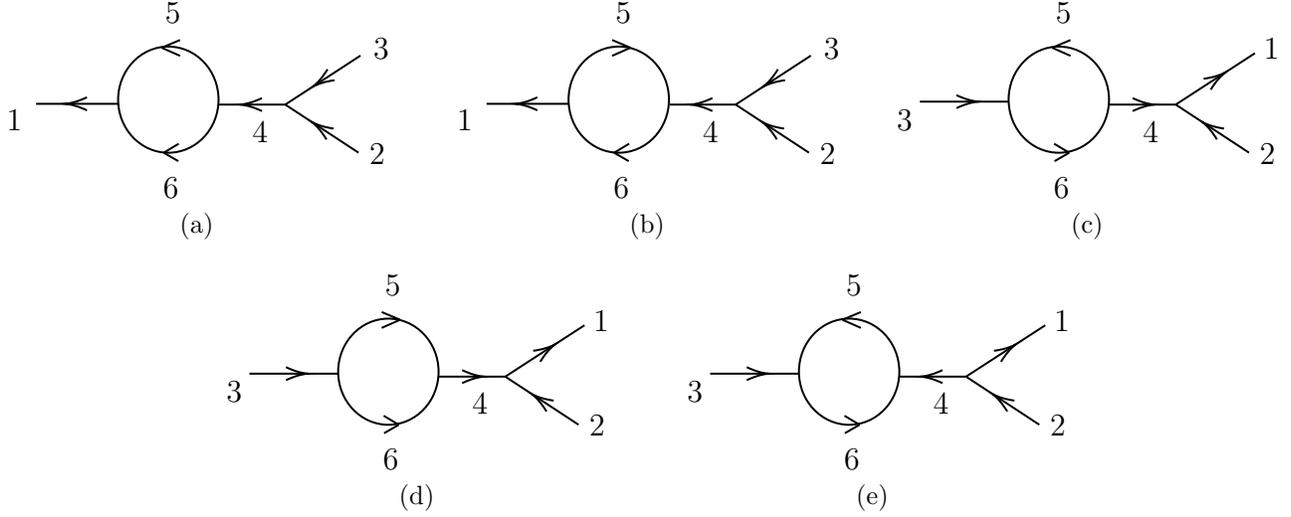

The leading order term $ \la a_{p_1} \da_{p_2} \da_{p_3}\ra_{\lam}$, given earlier in (\ref{319}), gives the standard (leading order) kinetic equation. The next-to-leading order term $\la a_{p_1} \da_{p_2} \da_{p_3}\ra_{\lam^3} $ gives the first corrections to this. The terms appearing here are: $V_a$, which arrises from the diagram shown earlier in Fig.~\ref{31loop}(a), $V_b$ which comes from Fig.~\ref{31loop}(a) but with the arrow on line $5$ reversed, $P_a$ which comes form the diagram shown earlier in Fig.~\ref{Threepillow}(a) and shown again in Fig.~\ref{PropRenorm}(a), and $P_b, P_c, P_d, P_e$ which have different choices of arrows on the propagator renormalization loop and different choices of lines which get the propagator renormalization. Explicitly, 
 \bml \nn
V_a =2 \sum_{p_4,p_5,p_6} \lam_{426} \lam_{356}^* \lam_{145} \prod_{i=1}^6 n_i  \frac{1}{\o_{p_2,p_3;p_1}{+}i\epsilon}\\
\Big\{ \frac{1}{n_1} \frac{1}{\o_{p_2,p_3;p_4, p_5}{+}i\eps}\Big[ \frac{1}{\o_{p_3;p_5,p_6}{+}i\eps} \Big(\frac{1}{n_4}{-}\frac{1}{n_2}\Big)\Big(\frac{1}{n_5} {+}\frac{1}{n_6}{-}\frac{1}{n_3}\Big) + \frac{1}{\o_{p_2,p_6;p_4}{+}i\eps} \Big(\frac{1}{n_5}{-}\frac{1}{n_3}\Big)\Big(\frac{1}{n_4} {-}\frac{1}{n_2}{-}\frac{1}{n_6}\Big) \Big] \\
 -\frac{1}{n_2} \frac{1}{\o_{p_3,p_4;p_1,p_6}{+}i\eps}\Big[ \frac{1}{\o_{p_3;p_5,p_6}{+}i\eps} \Big(\frac{1}{n_1}{-}\frac{1}{n_4}\Big)\Big(\frac{1}{n_5} {+}\frac{1}{n_6}{-}\frac{1}{n_3}\Big) + \frac{1}{\o_{p_4,p_5;p_1}{+}i\eps} \Big(\frac{1}{n_6}{-}\frac{1}{n_3}\Big)\Big(\frac{1}{n_1} {-}\frac{1}{n_4}{-}\frac{1}{n_5}\Big) \Big] \\
 -\frac{1}{n_3} \frac{1}{\o_{p_2,p_5,p_6;p_1}{+}i\eps}\Big[ \frac{1}{\o_{p_4,p_5;p_1}{+}i\eps} \Big(\frac{-1}{n_2}{-}\frac{1}{n_6}\Big)\Big(\frac{1}{n_1} {-}\frac{1}{n_4}{-}\frac{1}{n_5}\Big) + \frac{1}{\o_{p_2,p_6;p_4}{+}i\eps} \Big(\frac{1}{n_1}{-}\frac{1}{n_5}\Big)\Big(\frac{-1}{n_2} {+}\frac{1}{n_4}{-}\frac{1}{n_6}\Big) \Big] \Big\}~
\end{multline}
 \bml \nn
V_b =2\sum_{p_4,p_5,p_6} \lam_{426} \lam_{635} \lam_{451}^* \prod_{i=1}^6 n_i  \frac{1}{\o_{p_2,p_3;p_1}{+}i\epsilon}\\
\Big\{ \frac{1}{n_1} \frac{1}{\o_{p_2,p_3,p_5;p_4,}{+}i\eps}\Big[ \frac{1}{\o_{p_3,p_5;p_6}{+}i\eps} \Big(\frac{1}{n_4}{-}\frac{1}{n_2}\Big)\Big(\frac{-1}{n_5} {+}\frac{1}{n_6}{-}\frac{1}{n_3}\Big) + \frac{1}{\o_{p_2,p_6;p_4}{+}i\eps} \Big(\frac{-1}{n_5}{-}\frac{1}{n_3}\Big)\Big(\frac{1}{n_4} {-}\frac{1}{n_2}{-}\frac{1}{n_6}\Big) \Big] \\
 -\frac{1}{n_2} \frac{1}{\o_{p_3,p_4;p_1,p_6}{+}i\eps}\Big[ \frac{1}{\o_{p_3,p_5;p_6}{+}i\eps} \Big(\frac{1}{n_1}{-}\frac{1}{n_4}\Big)\Big(\frac{-1}{n_5} {+}\frac{1}{n_6}{-}\frac{1}{n_3}\Big) + \frac{1}{\o_{p_4;p_1.p_5}{+}i\eps} \Big(\frac{1}{n_6}{-}\frac{1}{n_3}\Big)\Big(\frac{1}{n_1} {-}\frac{1}{n_4}{+}\frac{1}{n_5}\Big) \Big] \\
 -\frac{1}{n_3} \frac{1}{\o_{p_2,p_6;p_1,p_5}{+}i\eps}\Big[ \frac{1}{\o_{p_4;p_1,p_5}{+}i\eps} \Big(\frac{-1}{n_2}{-}\frac{1}{n_6}\Big)\Big(\frac{1}{n_1} {-}\frac{1}{n_4}{+}\frac{1}{n_5}\Big) + \frac{1}{\o_{p_2,p_6;p_4}{+}i\eps} \Big(\frac{1}{n_1}{+}\frac{1}{n_5}\Big)\Big(\frac{-1}{n_2} {+}\frac{1}{n_4}{-}\frac{1}{n_6}\Big) \Big] \Big\}~
\end{multline}
\bml \nn
P_a=-\frac{1}{2}\sum_{p_5,p_6} \lam_{123}  |\lam_{156}|^2 \prod_{i=1}^3 n_i\, n_5 n_6 \frac{1}{\o_{p_2,p_3; p_1}{+}i\eps} \frac{1}{\o_{p_2, p_3;p_5,p_6}{+}i\eps}\Big[ \frac{1}{\o_{p_2,p_3; p_1}{+}i\eps}  \(\frac{1}{n_5}{+}\frac{1}{n_6}\)\(\frac{1}{n_2}{+}\frac{1}{n_3}{-}\frac{1}{n_1}\)\\
- \frac{1}{\o_{p_1;p_5,p_6}{+}i\eps}  \(\frac{1}{n_2}{+}\frac{1}{n_3}\)  \(\frac{1}{n_1}{-}\frac{1}{n_5}{-}\frac{1}{n_6}\)  \Big] 
\end{multline} 
\bml  \nn
P_b=-\sum_{p_5,p_6} \lam_{123}|\lam_{651}|^2 \prod_{i=1}^3 n_i\, n_5 n_6 \frac{1}{\o_{p_2,p_3; p_1}{+}i\eps}   \frac{1}{\o_{p_2, p_3, p_5;p_6}{+}i\eps}\Big[ \frac{1}{\o_{p_2,p_3; p_1}{+}i\eps}  \(\frac{-1}{n_5}{+}\frac{1}{n_6}\)\(\frac{1}{n_2}{+}\frac{1}{n_3}{-}\frac{1}{n_1}\)\\
- \frac{1}{\o_{p_1,p_5;p_6}{+}i\eps}  \(\frac{1}{n_2}{+}\frac{1}{n_3}\)  \(\frac{1}{n_1}{+}\frac{1}{n_5}{-}\frac{1}{n_6}\) \Big]
\end{multline} 
\bml \nn
P_c=2\sum_{p_5,p_6} \lam_{123}  |\lam_{653}|^2 \prod_{i=1}^3 n_i\, n_5 n_6 \frac{1}{\o_{p_2,p_3; p_1}{+}i\eps} \frac{1}{\o_{p_2, p_6;p_1,p_5}{+}i\eps}\Big[ \frac{1}{\o_{p_2,p_3; p_1}{+}i\eps}  \(\frac{1}{n_5}{-}\frac{1}{n_6}\)\(\frac{1}{n_2}{+}\frac{1}{n_3}{-}\frac{1}{n_1}\)\\
- \frac{1}{\o_{p_6;p_3,p_5}{+}i\eps}  \(\frac{1}{n_2}{-}\frac{1}{n_1}\)  \(\frac{-1}{n_3}{-}\frac{1}{n_5}{+}\frac{1}{n_6}\)  \Big] 
\end{multline} 
\bml \nn
P_d=\sum_{p_5,p_6} \lam_{123}  |\lam_{356}|^2 \prod_{i=1}^3 n_i\, n_5 n_6 \frac{1}{\o_{p_2,p_3; p_1}{+}i\eps} \frac{1}{\o_{p_2, p_5,p_6;p_1}{+}i\eps}\Big[ \frac{1}{\o_{p_2,p_3; p_1}{+}i\eps}  \(\frac{-1}{n_5}{-}\frac{1}{n_6}\)\(\frac{1}{n_2}{+}\frac{1}{n_3}{-}\frac{1}{n_1}\)\\
- \frac{1}{\o_{p_5,p_6;p_3}{+}i\eps}  \(\frac{1}{n_2}{-}\frac{1}{n_1}\)  \(\frac{-1}{n_3}{+}\frac{1}{n_5}{+}\frac{1}{n_6}\)  \Big] 
\end{multline} 
\bml \nn
P_e =-2\sum_{p_5,p_6} \lam_{213}^* \lam_{536}\lam_{635}\prod_{i=1}^3 n_i\, n_5 n_6\, \frac{1}{\o_{p_2,p_3; p_1}{+}i\eps}\Big\{- \frac{1}{\o_{p_2;p_1, p_3}{+}i\eps} \frac{1}{\o_{p_2, p_6; p_1,p_5}{+}i\eps} \(\frac{1}{n_5}{-}\frac{1}{n_6}\)\(\frac{1}{n_2}{-}\frac{1}{n_1}{-}\frac{1}{n_3}\)\\
+ \frac{1}{\o_{p_3,p_6;p_5}{+}i\eps} \(  \frac{1}{\o_{p_2, p_6;p_1,p_5}{+}i\eps}+ \frac{1}{2\o_{p_3}{+}i\eps}\) \(\frac{1}{n_2}{-}\frac{1}{n_1}\)  \(\frac{1}{n_3}{-}\frac{1}{n_5}{+}\frac{1}{n_6}\)\\
+\frac{1}{\o_{p_3, p_5;p_6}{+}i\eps}\frac{1}{2\o_{p_3}{+}i\eps}   \(\frac{1}{n_2}{-}\frac{1}{n_1}\)  \(\frac{1}{n_5}{-}\frac{1}{n_6}{+}\frac{1}{n_3}\)
 \Big\}~.
\end{multline}
Further discussion of terms in the kinetic equation is given in Appendix~\ref{apc}.

\section{Prescription for a general Feynman diagram} \label{sec32}

In this section we give an algorithm for computing the contribution of any Feynman diagram to an equal-time correlation function. In Sec.~\ref{sec2} we already showed that correlation functions can be computed order by order in the nonlinear interaction ($\lam$), just as one does in quantum field theory. The remaining task is, for a given Feynman diagram, to perform the integration over the intermediate times (or, equivalently, frequencies) appearing in the loops. We did this explicitly in the previous section, for some particular one loop diagrams. From the several diagrams that we evaluated, we can deduce the rule for a general diagram. We first state the rules and then derive them.\\

\noindent \textbf{Rules:}

\begin{enumerate}
\item Pick an ordering of the times at each vertex. The time $t_a$ at which the correlation function is being evaluated must be the smallest time.
\item Start at the latest time on the diagram and move from vertex to vertex in decreasing order of their times, until finally reaching the vertex at the earliest time. Each next vertex must be a neighbor of at least one previously visited vertex. 
\item At each step in this process, draw an imaginary loop  enclosing all vertices visited so far. Write down a factor of 
\be \label{step3}
\frac{-1}{\o_{p_i} {+}\o_{p_j} {+}\ldots {-} \o_{p_a} {-} \o_{p_b}{-} \ldots {+} i\eps}~,
\ee
where $\o_i, \o_j, \ldots$ are the frequencies of all lines entering this imaginary loop and $\o_a, \o_b, \ldots$ are the frequencies of all lines leaving the imaginary loop.

\item In addition, at each step in the process, write down a factor of 
\be \label{step4}
\(\frac{1}{n_k} {+}\frac{1}{n_l} {+}\ldots {-}\frac{1}{n_{\al}} {-}\frac{1}{n_{\beta}} {-}\ldots \)~,
\ee
where $\o_k, \o_l, \ldots$ are the frequencies of the lines entering the vertex and $\o_{\al}, \o_{\beta}$ are the frequencies of the lines leaving the vertex. However, omit an $\frac{1}{n_i}$ if it has already been included at a previous step. 
\item Multiply the resulting expression by a product of $n_i$, one for each frequency $\o_i$ appearing in the diagram, and multiply by a product of the couplings for each vertex, and sum the resulting expression over all internal momenta. 
\item Repeat Step 1 through 5 for all possible time orderings and sum the results. Include the appropriate Feynman diagram combinatorics factor. 
\end{enumerate}

\noindent \textbf{Justification:}
Let us understand these rules. Suppose for concreteness that we have four vertices with the time ordering $t_d{>}t_c{>}t_b{>}t_a$. The integrals we need to do are, 
\be
\int_{t_a}^{\infty} d t_b\, e^{- i\sigma_b t_b} \int_{t_b}^{\infty} d t_c\, e^{ -i \sigma_c t_c} \int_{t_c}^{\infty} d t_d\, e^{- i \sigma_d t_d} =\frac{-i}{(\sigma_b{+}\sigma_c{+}\sigma_d)}\frac{-i}{(\sigma_c{+} \sigma_d)} \frac{-i}{\sigma_d}~.
\ee
Here the $\sigma$ are the sum of the outgoing minus ingoing frequencies from a vertex; we get these through the delta functions we insert, see below (\ref{321}). One now sees that as we successively do the integrals, starting from the latest time, we get the rules listed in Steps 2 and 3. Note that all our integrals here converge (because  the relevant poles of the $\o$ integrals  were chosen to ensure this), so the contributions from infinite time vanish. 

Now let us understand the vertex factor, appearing in Step 4. According to the Feynman rules, each vertex comes with a sum of $g_i^*$ for each  propagator leaving the vertex and a $- g_i$ for each propagator  entering the vertex, see (\ref{vertexRule}). At the poles, $\o_i = \o_{p_i}\pm \g_i$, we see from (\ref{gi}) that $g_i$ and $g_i^*$ become, 
\be
g_i \rightarrow   \begin{cases} \frac{1}{n_i}~, \ \ \ \o_i = \o_{p_i} {+} i\g_i \\ 
0 ~, \ \ \ \  \o_i = \o_{p_i} {-}i\g_i  \end{cases}~ \ \ \ \ g_i^* \rightarrow   \begin{cases}0~, \ \ \ \ \o_i = \o_{p_i} {+}i \g_i \\ 
 \frac{1}{n_i} ~, \ \ \  \o_i = \o_{p_i} {-}i\g_i  \end{cases}~.
\ee
So from each vertex we will get a sum and difference of various $1/n_i$. We need only determine if for a propagator entering or leaving a vertex we get a $1/n_i$ or nothing. Consider two vertices next to each other, as in Fig.~\ref{41loop}(a). In the corresponding vertex factor (\ref{321}), either the $g_5{+}g_6$ term coming from vertex $b$ gives $1/n_5 {+}1/n_6$ and the  $g_5^*{+}g_6^*$ term coming from vertex $c$ gives $0$, or vice-versa. Which option is the correct one depends on the time ordering of $t_b$ and $t_c$, which determines if the $\o_5, \o_6$ integrals are closed in the upper or lower half plane. In particular, the vertex at the later time gets to have the $1/n_5 {+}1/n_6$. Since in our prescription we proceed from later to earlier time, this explains Step 4.

\begin{figure}[h]
\centering
\subfloat[]{
\includegraphics[width=1.5in]{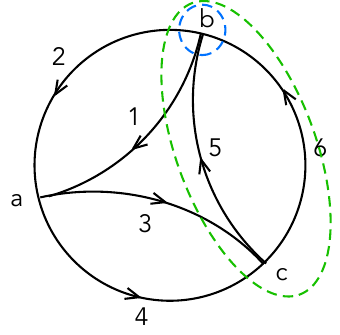}}
 \ \ \ \  \ \ \ \ \ \ \ \  \  \ \ \ \ 
\subfloat[]{
\includegraphics[width=1.5in]{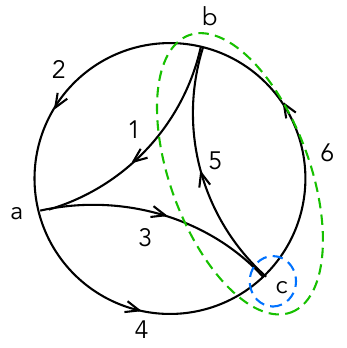}
} \ \ \ \ \ \ \ \ \ \ \ \ 
\subfloat[]{
\includegraphics[width=1.8in]{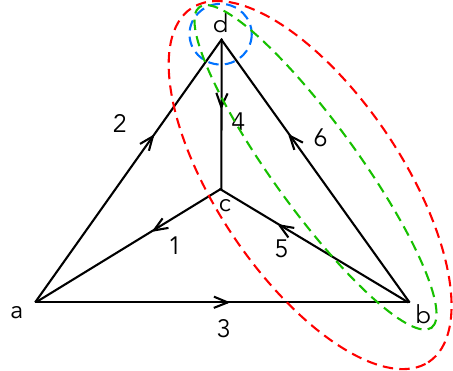}
}
\caption{Applying the rules to four-point and three-point correlation functions.} \label{looprules}
\end{figure}

\vspace{.3cm}
 \noindent \textbf{Examples:}
Let us look at some examples of applying these rules. Consider again the one-loop diagram for the quartic interaction, Fig.~\ref{41loop}. Take the time ordering $t_b{>}t_c{>}t_a$. We draw a loop around vertex $b$, shown in blue in Fig.~\ref{looprules}(a). Step 3 instructs us to write a factor of 
\be
\frac{-1}{\o_{p_5,p_6;p_1,p_2}{+}i\eps}~,
\ee
and Step 4 instructs us to write  a factor of 
\be \label{335}
\(\frac{1}{n_5}{+}\frac{1}{n_6}{-}\frac{1}{n_1}{-}\frac{1}{n_2}\)~.
\ee
The next largest time is $t_c$, so we expand our loop (shown in green) to include vertex $c$. Step 3 instructs us to write a factor of
\be
\frac{-1}{\o_{p_3,p_4;p_1,p_2}{+}i\eps}~,
\ee
and Step 4 instructs us to write  a factor of 
\be
\(\frac{1}{n_3}{+}\frac{1}{n_4}\)~.
\ee
As Step 4 says, even though $\o_1$ and $\o_2$ are leaving the green circle, we do not include $-\frac{1}{n_1}{-}\frac{1}{n_2}$ since this term already appeared earlier, when accounting for the blue loop around vertex $b$, (\ref{335}). Implementing Step 5, we reproduce (\ref{325}). 
The other time ordering, $t_c{>}t_b{>}t_a$, proceeds in a similar fashion. We have drawn the corresponding loops in Fig.~\ref{looprules}(b). 

Let us  now look again at the loop digram for the cubic interaction, Fig.~\ref{31loop}, which we evaluated earlier. There are six possible time orderings. Let us look at, for instance, the one in which $t_d{>}t_b{>}t_c{>}t_a$. We draw a loop, shown in green in Fig.~\ref{looprules}(c), around vertex $d$, which is at the latest time. Frequencies $\o_2$ and $\o_6$ are entering and $\o_4$ is leaving. Steps 3 and 4 instruct us to write a factor of 
\be \label{338}
\frac{-1}{\o_{p_2,p_6;p_4}{+}i\eps} \(\frac{1}{n_2}{+}\frac{1}{n_6}{-}\frac{1}{n_4}\)~.
\ee
Expanding the loop to include the  vertex $b$, which is at the next largest time, we get the loop shown in green in  Fig.~\ref{looprules}(c). Frequencies $\o_2$ and $\o_3$ are entering and $\o_4$ and $\o_5$ are leaving. Steps 3 and 4 instruct us to write a factor of 
\be \label{339}
\frac{-1}{\o_{p_2,p_3;p_4,p_5}{+}i\eps} \(\frac{1}{n_3}{-}\frac{1}{n_5}\)~, 
\ee
where there is no $1/n_2{-} 1/n_4$, since it was already included in (\ref{338}).  Finally, we expand the loop to the loop shown in red in Fig.~\ref{looprules}(c), so as to include  vertex $c$.  Frequencies $\o_2$ and $\o_3$ are entering and $\o_1$ is leaving. Steps 3 and 4 instruct us to write a factor of 
\be \label{340}
\frac{-1}{\o_{p_2,p_3;p_1}{+}i\eps} \frac{1}{n_1}~, 
\ee
where there is no $1/n_2$ or $1/n_3$, since both were already included. 
Multiplying (\ref{338}), (\ref{339}), and (\ref{340}), we reproduce the second of the six terms in (\ref{329}).

\begin{figure}[h]
\centering
\subfloat[]{
\includegraphics[width=1.7in]{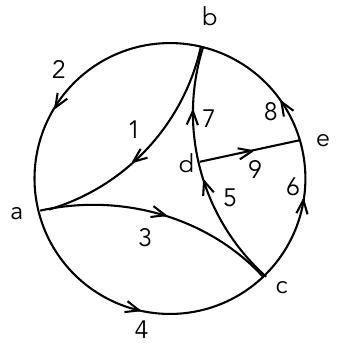}} \ \ \ \ \ \ \ \ \ \ 
\subfloat[]{
\includegraphics[width=1.7in]{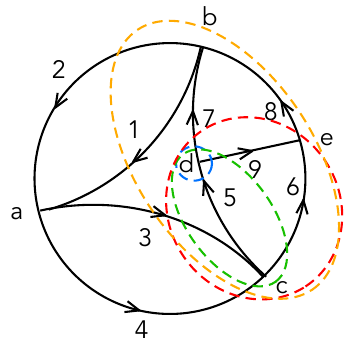}}
\caption{ (a) An example of a higher order diagram. (b) Applying our rules to a particular time ordering, resulting in the contribution (\ref{lll}).} \label{cubicquartic}
\end{figure}

As our final example, let us look at an example of a diagram that occurs if we have both cubic and quartic interactions. One such diagram (chosen for no particular reason) is shown in Fig.~\ref{cubicquartic} (a). It contributes to the equal time four-point function $\la a_1 a_2 \da_3 \da_4\ra (t_a)$.  There are many different time orderings one needs to consider. To illustrate, let us pick one  such ordering:
$t_d{>}t_c{>}t_e{>}t_b{>}t_a$. Applying the rules we get the contribution, 
\bml \label{lll}
\sum_{p_i} \lam_{1278}\lam_{5634}\lam_{869}\lam_{579}^*\prod_{i=1}^9 n_i\, \frac{-1}{\o_{p_5;p_7,p_9}{+}i\eps} \(\frac{1}{n_5}{-}\frac{1}{n_7}{-}\frac{1}{n_9}\)\frac{-1}{\o_{p_3,p_4;p_6,p_7,p_9}{+}i\eps} \(\frac{1}{n_3}{+}\frac{1}{n_4}{-}\frac{1}{n_6}\)\\
\frac{-1}{\o_{p_3,p_4;p_7,p_8}{+}i\eps} \({-}\frac{1}{n_8}\)\frac{-1}{\o_{p_3,p_4;p_1,p_2}{+}i\eps} \({-}\frac{1}{n_1}{-}\frac{1}{n_2}\)~,
\end{multline}
where reading from left to right, the terms correspond to larger and larger loops shown in Fig.~\ref{cubicquartic} (b).

\section{Discussion} \label{sec6}

In \cite{RS1} we gave a systematic method for computing higher order terms in the wave kinetic equation: correlation functions are defined by averaging over Gaussian random forcing, which is set to zero at the end. This stochastic classical field theory is equivalent to a quantum field theory. One computes equal-time correlation functions in the quantum field theory, order by order in the number of interaction vertices, and inserts into the right hand side of (\ref{220}) to get the kinetic equation. The only nontrivial step is evaluating the Feynman diagrams with loops, which involve  integrals over the intermediate times (or frequencies) inside the loops. In this paper we have shown how to carry out the integrals in general, obtaining a simple prescription for writing the result, described in Sec.~\ref{sec32}. 

Our answer still has integrals over the loop momenta; these  need to be done on a case by case basis, as they are functions of the couplings, $\lam_{p_1 p_2 p_3}$ and $\lam_{p_1 p_2 p_3 p_4}$, which are momentum dependent and system dependent. If these loop integrals over momenta didn't have UV or IR divergences, we would be close to being done --- the corrections to the leading order kinetic equation would always be small, and the largest corrections would come from the loop diagrams with the fewest number of interaction vertices. But we're not done: the loops often have  UV and IR divergences \cite{FR}. This introduces additional small and large parameters which compete with the smallness of the nonlinearity parameter, and make the perturbation theory multiscale. We must decide which classes of loop diagrams are dominant and sum those. In other words, we must renormalize. The results in this paper provide the necessary tools to  carry out this task, which will be the subject of future work.

\sss*{Acknowledgments} 
We thank G.~Falkovich for many discussions and collaboration on related work. 
This work  is supported in part by NSF grant PHY-2209116, by BSF grant 2022113, and by the ITS through a Simons grant. 
The work of MS is supported by the  Israeli Science Foundation Center of Excellence (grant No.~2289/18). VR and MS thank the Aspen Center for Physics (NSF grant  PHY-2210452) where part of this work was completed. 

\appendix

\section{Perturbative Liouville equation for waves} \label{sec:Liouv}

An alternate method for studying wave turbulence is to average over phase space, instead of introducing external forcing and dissipation. In particular, one aims to find the evolution of the phase space density $\rho(J_i,\al_i,  t)$, a function of the action, $J_i$, and angle, $\al_i$, variables, with $i$ running over the different modes of the field. This method was used in \cite{RS2} to find the next-to-leading order correction to the kinetic equation for a theory with four-wave interactions. Here we extend this to the theory with three-wave interactions, like the one studied in the main body, reproducing the results found there. In fact, we go further: we show that at every order in perturbation theory, the results of this method agree with the one in the main body of the text.~\footnote{ An alternate way of seeing the equivalence, which will be discussed in \cite{Schubring}, is to solve for the complete evolution of the phase space density when one has random forcing (the Fokker-Planck equation), and then either first average over forcing and then phase space, or vice-versa.}

\subsection*{Perturbative solution of the Liouville equation} \label{A1}
The evolution of the phase space density is governed by the Liouville equation \cite{Prigogine}, 
\be \label{Liou1}
i \frac{\d \rho}{\d t} = \(L_0 + \delta L\) \rho~,
\ee
\be \label{B2}
L_0 =  - i \v \o \cdot\frac{\d }{\d \v \al}~,\ \ \ \ \ \ \ \ \ \ \ \delta L= i\( \frac{\d H_{int}}{\d \v\al}\cdot \frac{\d}{\d {\v J}} -\frac{\d H_{int}}{\d\v J}\cdot \frac{\d}{\d \v\al} \)~,
\ee
where $L_0$ is due to the free part of the Hamiltonian, $H_2$ (which only depends on the action variables), and $\delta L$ is due to the interacting part, $H_{int}$. We have defined, $ \o_j \equiv \frac{\d H_2}{\d J_j}$. 
  
The solution for a time-independent phase space density is given by iterating, 
\be \label{B3}
\la \v n | \Phi\ra = \la \v n | \Phi_0\ra + \sum_{\v n'}G(\v n)  \la \v n| \delta L |\v n'\ra \la\v n' |\Phi\ra~,
\ee
where the eigenfunctions $|\v n\ra$ of the free Liouville operator $L_0$ are plane waves $\la \v\al| \v n\ra = \exp\( i \v n \cdot \v \a\)$ and $G(\v n)$ is the propagator,
\be \label{Gn}
G(\vec n) = \frac{1}{-\vec n \cdot \vec \o +i\eps}~.
\ee
The kinetic equation is then given by, 
  \be \label{kep}
 \frac{\d n_r}{\d t} \propto \int dJ\, J_r\, \delta \mL~, \ \ \ \ \text{where}  \ \ \delta \mL =  \sum_{\v n' } \la \v 0| \delta L |\v n'\ra \la \v n' |\Phi\ra~.
 \ee
 Inserting (\ref{B3}) gives $ \delta \mL =\(\delta \mathcal{L}\)_{\text{first}} + \(\delta \mathcal{L}\)_{\text{second}} + \ldots $, where
 \bea \label{B5}
 \(\delta \mL \)_{\text{first}} =& &
 \sum_{\v n'}\la \v 0| \delta L |\v n'\ra G(\v n ')\la \v n'| \delta L |0\ra \rho(J)\\ \label{B6}
 \(\delta \mathcal{L}\)_{\text{second}}&=&\sum_{\v n', \v n''}  \la \v 0| \delta L |\v n'\ra G(\v n ')\la \v n'| \delta L|\v n''\ra G(\v n'')  \la \v n'' |\delta L   |0\ra \rho(J) + \ldots
\eea
where $\rho(J) \equiv \la 0 |\Phi_0\ra $. We will take $\rho(J)$ to be a Gaussian, 
 \be  \label{rhoJg}
 \rho (J)= \frac{1}{\prod_i n_i} \exp\( - \sum_i \frac{J_i}{n_i}\)~, \ \  \ \ \ \ \la J_i\ra = n_i~, \ \ \ 
 \ee 
 
The main point is this: the phase space integrated over all angles, $\rho(J)$, which has no angular dependence, is governed by a differential equation, the right-hand side of which encodes a series of transitions to intermediate states which have angular dependence. In particular, each interaction vertex gives $q$ of the different modes angular dependence (for a $q$-body interaction). We start and end in a state with no angular dependence.

  In \cite{RS2} we specialized these equations to interacting waves with a quartic interaction. Here we look at these equations for a field theory with a cubic interaction. 
  The Hamiltonian, $H_2 + H_3$, takes the following form when written in action-angle variables, $ a_p = \sqrt{J_p} e^{-i \al_p}$, 
\be\label{HcubicJ}
H =\sum_p \o_p J_p +  \frac{1}{2}\sum_{p_i}\sqrt{J_1 J_2 J_3}\( \lam_{123} e^{i (\al_1 {-} \al_2 {-} \al_3)}+\text{c.c}\)~.
\ee
Computing the Liouville operator we have that $L_0$ is (\ref{B2}) with $\omega_i$ given by $\omega_p$  and $\delta L$ in (\ref{B2}) is equal to, 
\be \label{B12}
\delta L =  \frac{1}{2}\sum_{1,2,3} (-\lam_{123}) \sqrt{J_1 J_2 J_3} e^{- i \v e_{1; 2,3} \cdot \v \al} \[ \(\d_1 -\d_2- \d_3 \)+ \frac{i}{2}\(\frac{\d_{\al_1}}{J_1}+ \frac{\d_{\al_2}}{J_2}+\frac{\d_{\al_3}}{J_3}\)\] - \text{c.c.}~,
\ee
where $\d_i \equiv \d_{J_i}$ and where  $\v e_{i; j,k} $, which is a vector that has a $-1$ in the $i$'th entry and a $+1$ in the $j$'th and $k$'th, and a zero elsewhere; in other words, $-\v e_{i; j,k} \cdot \v \al = \al_i{-}\al_j{-}\al_k$. The matrix element of $\delta L$ trivially follows, 
\bea  \nn
\!\!\!\!\!\!\! \la \v n| \delta L |\v n'\ra &=& -  \frac{1}{2}\sum_{1,2,3} \lam_{123} \sqrt{J_1 J_2 J_3} \, \delta_{\v n + \v e_{1;2,3}, \v n'} \[ \(\d_1  -\d_2- \d_3 \)-\frac{1}{2}\(\frac{n'_1}{J_1}+ \frac{n'_2}{J_2}+\frac{n'_3}{J_3} \)\]\\
&&- \frac{1}{2}\sum_{1,2,3} \lam^*_{123} \sqrt{J_1 J_2 J_3} \, \delta_{\v n +\v e_{2,3;1}, \v n'} \[ \(\d_2  +\d_3- \d_1 \)-\frac{1}{2}\(\frac{n'_1}{J_1}+\frac{n'_2}{J_2}+\frac{n'_3}{J_3} \)\]
~, \label{Lmatc}
\eea
where $\v e_{2,3;1} = - \v e_{1;2,3}$. 

\subsection*{Leading order} \label{A2}
\begin{figure}
\centering
\subfloat[]{
\includegraphics[width=1.3in]{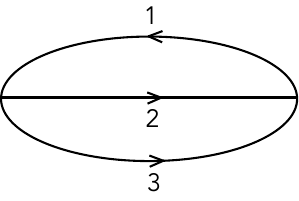}} \ \ \ \  \ \ \ \ \ \ \ \  \  \ \ \ \ 
\subfloat[]{
\includegraphics[width=1.3in]{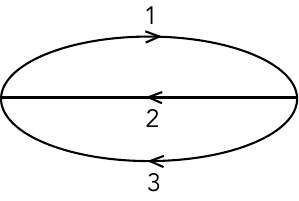}}
\caption{ Intermediate states (a) $|\b1; 23\ra$, and (b) $|\b2 \b3; 1\ra$. Our notation is such that when we write $| \b 1;  23\ra$ we mean a state in which mode $1$ has occupation number $-1$ and modes $2$ and $3$ have occupation number $1$.   } \label{PtreeC}
\end{figure}
We now start iteratively computing the kinetic equation (\ref{kep}), starting with  $\(\delta \mL \)_{\text{first}}$ in (\ref{B5}). There are two diagrams, as shown in Fig.~\ref{PtreeC}. The first, Fig.~\ref{PtreeC}(a),  gives the following contribution to $ \(\delta \mathcal{L}\)_{\text{first}}$,
\be \label{58}
\sum_{1,2,3} G(\v e_{1;23}) \la \v 0| \delta L| \b 1;  23\ra\la \b 1; 23|\delta L |\v 0\ra  \rho(J)~, \ \ \ \   \text{where }\ \ \ \ \ G(\v e_{1;23}) = \frac{1}{\o_{p_1; p_2, p_3}{+}i\eps}~,
\ee
which, upon evaluating, gives, 
\be
\frac{1}{4} \sum_{1,2,3} |\lam_{123}|^2 \frac{1}{\o_{p_1;p_2,p_3}{+}i\eps}  \(\d_1 {-}\d_2{-} \d_3 \) J_1 J_2 J_3 \(\d_2  {+}\d_3{-} \d_1 \) \rho(J)~,
\ee
where we made use of the commutator  $\[\d, \sqrt{J}\] =\frac{1}{2J} \sqrt{J}$. 
The second diagram is just the first diagram with arrows reversed, which gives the same contribution but with a denominator that has an extra minus sign for $\o_{p_1;p_2,p_3}$. Adding the two and using (\ref{kep}) we recover the leading order kinetic equation,
\be
\frac{\d n_a}{\d t} = 2\pi\sum_{1,2,3} |\lam_{123}|^2 (\delta_{1a} {-}\delta_{2a} {-}\delta_{3a}) \delta(\o_{p_1;p_2 p_3})  \(\frac{1}{n_1} {-}\frac{1}{n_2}{-}\frac{1}{n_3}\) n_1 n_2 n_3~.
\ee

\subsection*{Next-to-leading order} \label{A3}
We now look at the terms that appear at next order. All terms in $(\delta \mL)_{\text{second}}$ (\ref{B6}), with two intermediate states vanish, so we look at $(\delta \mL)_{\text{third}}$, which has three intermediate states. 
\begin{figure}[h]
\centering
\includegraphics[width=1.6in]{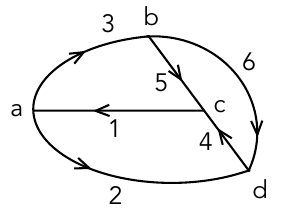}
\caption{ We start with the vacuum. The intermediate states are then $|\b 1; 23\ra$, followed by $|\b 1; 2 5 6\ra$, followed by $|\b 4; 26\ra$, and ending in the vaccum.}  \label{cubic1loopArrow0}
\end{figure}
One such sequence of intermediate states is shown in Fig~\ref{cubic1loopArrow0}. Evaluating the contribution of this diagram gives, 
\bml \label{T1}
\delta \mL_{t} (\{a,b,c,d\})=\sum_{1,\ldots, 6} \lam_{123}\lam_{426}^* \lam_{356} \lam_{145}^*G(\v e_{1;23}) G(\v e_{1; 256}) G(\v e_{4;26}) \\
\(\d_1{-}\d_2 {-}\d_3\) J_1 J_2 J_3\(\d_3 {-} \d_5{-}\d_6\) J_5 J_6 \(\d_4 {+}\d_5 {-}\d_1\) J_4  \(\d_2{+}\d_6{-}\d_4\) \rho(\v J)~.
\end{multline}
Using $\rho(\v J)$ (\ref{rhoJg}) and evaluating the integral of $\delta \mL_t( \{a,b,c,d\})$ multiplied by $J_r$, as prescribed by (\ref{kep}),  gives, 
\bml \label{Jrabcd1}
\int d J\, J_r\, \delta \mL_t( \{a,b,c,d\}) 
= \sum_{1,\ldots, 6} \lam_{123}\lam_{426}^* \lam_{356} \lam_{145}^*G(\v e_{1;23}) G(\v e_{1; 256}) G(\v e_{4;26}) \\
\(\delta_{1r} {-}\delta_{2r} {-}\delta_{3r}\) \frac{1}{n_3}\Big(\frac{1}{n_1}{-}\frac{1}{n_5}\Big)\Big(\frac{1}{n_2} {+}\frac{1}{n_6}{-}\frac{1}{n_4}\Big)\prod_{i=1}^6 n_i~.
\end{multline}
This matches the last term in (\ref{329}). In more detail, in the main body of the text, the kinetic equation resulted by inserting contributions to the equal-time three-point function, such as (\ref{329}), into (\ref{220}). The contribution to the kinetic equation of the last term of (\ref{329}) is what essentially matches (\ref{Jrabcd1}). Notice that (\ref{220}) has a factor of $(\delta_{kp_1}{-}\delta_{kp_2}{-}\delta_{kp_3})$ in front, which is just the factor of $\(\delta_{1r} {-}\delta_{2r} {-}\delta_{3r}\)$ in (\ref{Jrabcd1}) in slightly different notation. 

There are other possible intermediate states. We can obtain them by changing the order in which the vertices appear. 
For instance, the ordering $a$, $d$, $c$, $b$, shown below, 
\begin{figure}[h]
\centering
\includegraphics[width=1.6in]{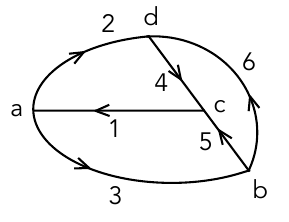}
\caption{ A different sequence of intermediate states: $|\b1; 23\ra$, followed by $|\b 1 \b 6; 34\ra$, followed by $|\b 5 \b6; 3\ra$.  }  \label{cubic1loopArrow}
\end{figure}
gives the contribution, 
\bml \label{T2}
\delta \mL_t(\{a,d,c,b\})=\sum_{1,\ldots, 6} \lam_{123}\lam_{426}^* \lam_{356} \lam_{145}^*G(\v e_{1;23}) G(\v e_{16; 34}) G(\v e_{56;3}) \\
\(\d_1{-}\d_2 {-}\d_3\) J_1 J_2 J_3 \(\d_2{+}\d_6{-}\d_4\) J_4 J_6 \(\d_4 {+}\d_5 {-}\d_1\) J_5 \(\d_3 {-} \d_5{-}\d_6\) \rho(\v J)~.
\end{multline}

Of course, both these diagrams are simply redrawings of the tetrahedron diagram that appeared in Fig.~\ref{31loop}. We need to consider all $24$ permutations of the vertices. 
The six terms with vertex $a$ first will give the six terms in (\ref{329}). The six terms with vertex $c$ first will give the same thing, but complex conjugated and with a relative minus sign. So their sum is just the imaginary part of the terms with vertex $a$ first. We therefore reproduce the contribution of (\ref{329}) to the kinetic equation (notice that (\ref{220}) takes the imaginary part of the coupling multiplying the three-point function). The terms with vertex $b$ first combined with the terms with vertex $d$ first (which are the complex conjugate), reproduce the contribution of the tetrahedron in Fig.~\ref{31loop}) with $5\rightarrow-5$. We may do a similar analysis for the propagator renormalization diagram, but this is unnecessary, as we will now give a general argument that the results will always match what is in the main body of the text.

\subsection*{General Prescription} \label{A4}

In implementing the perturbative solution of the Liouville equation, (\ref{B3}), we need to construct the sequences of all possible intermediate states. This can be represented by vacuum Feynman diagrams. Each Feynman diagram is topologically distinct. For each Feynman diagram we move around the vertices, so as to consider all possible orderings and correspondingly all possible intermediate states.  Without loss of generality, we can pick one of the vertices to be first, while considering different orderings of the other vertices. If we cut the Feynman diagram at this first vertex, separating the edges entering it, then we have a diagram that contributes to a $q$-point correlation (for a $q$-body interaction). In particular, it may be a diagram for either a correlator such as $\la a_1 \da_2 \da_2\ra$ (for $q=3$), or its complex conjugate $\la \da_1 a_2 a_3\ra$. The complex conjugate arrises if we reverse all the arrows on the Feynman diagram. We can account for this by including only one Feynman diagram out of each pair that have all arrows reversed relative to each other and then taking the imaginary part, as we saw earlier. We can  achieve this by drawing all diagrams contributing to $\la a_1 \da_2 \da_2\ra$ (for $q=3$)  and connecting the external lines together  to meet at a vertex. We then consider all possible orderings of the other vertices, which determines the intermediate states. 

At this stage the number of terms we have matches what was discussed in Sec.~\ref{sec32}. Now let us show that the terms are in fact the same. We get from vertex to vertex with a propagator
$\frac{1}{-\v n{\cdot} \v \o {+}i\eps}$, see (\ref{B3}), where $|\v n\ra$ is the current state. This matches the prescription given in Step 3 (\ref{step3}) of having a denominator with a sum of all incoming frequencies minus all outgoing frequencies. In our situation here the imaginary loop encloses all vertices visited so far. At each vertex we also have a transition matrix of $\delta L$, see (\ref{B12}) or (\ref{Lmatc}). We see that it it involves $\( \d_{k}{+}\d_l{+}\ldots {-}\d_{\al} {-}\d_{\beta} {-}\ldots\)$ where $\o_k, \o_l, \ldots$ are the frequencies of the lines entering the vertex and $\o_{\al}, \o_{\beta}, \ldots$ are the frequencies of the lines leaving the vertex. The transition matrix element also has factors of $\sqrt{J_i}$ and  of $n_i'/J_i$, see (\ref{Lmatc}). However, the latter can be eliminated by appropriately commuting through the $\sqrt{J_i}$. The end result, as seen in, for example, (\ref{T1}), is that one has each $J_i$ appear once, immediately to the right of when the corresponding $\d_i$ appears. After inserting $\rho(J)$ that is an exponential, (\ref{rhoJg}), it is clear we reproduce Steps 4 and 5 (\ref{step4}). Similar rules were found in \cite{Gurarie95, Polyakov}. 

\section{Symmetries} \label{sec:sym}
In this Appendix we show that the symmetries of the Hamiltonian, combined with the symmetries of the some of the Feynman diagrams, provide an efficient way of finding many of the contributions to the kinetic equation. We will, for instance, show that once one has any of the six terms appearing in the contribution of the tetrahedron diagram (\ref{329}) the five others follow by a symmetry transformation.

\subsection*{Symmetries of the Hamiltonian}
We first note the symmetries of the Hamiltonian, written in action-angle variables (\ref{HcubicJ}). The Hamiltonian is invariant under, 
\be \label{Hsym1}
\o_i \rightarrow -\o_i~, \ \ \ J_i \rightarrow - J_i~, \ \ \ \al_i \rightarrow  - \al_i~, \ \ \ \lam_{123} \rightarrow -i \lam_{-1 -2 -3} \equiv -i\lam_{123}^*~, \ \ \ \ i =1,2,3
\ee
Note that $\lam_{123}^*$ transforms in the same way, $\lam_{123}^* \rightarrow - i \lam_{123}$. 
The Hamiltonian is also invariant under a flip in sign of just two indices, 
\be
\o_i \rightarrow -\o_i~, \ \ \ J_i \rightarrow - J_i~, \ \ \ \al_i \rightarrow  - \al_i~, \ \ \ \lam_{123} \rightarrow  \lam_{-1 -2 3} \equiv \lam_{213}~, \ \ \ \ i =1,2
\ee
or a flip in one index,
\be
\o_2 \rightarrow -\o_2~, \ \ \ J_2 \rightarrow - J_2~, \ \ \ \al_2 \rightarrow  - \al_2~, \ \ \ \lam_{123} \rightarrow-i \lam_{1 -2 3} \equiv -i \lam^*_{321}~,.
\ee

This is useful because we sometimes have  diagrams which have some arrows flipped relative to other diagrams. Flipping an arrow means flipping the sign of $\alpha$. Due to the symmetries of the Hamiltonian we can instead flip the signs of the $\o_{p_i}$ and $J_{i}$, and change the couplings, as stated above. Note that in the kinetic equation we have $n_i$ instead of $J_i$ (the former is the expectation value of the latter). So an arrow going in the opposite direction on line $i$ means that in the contribution to the kinetic equation  $n_i$, $\o_{p_i}$, and $\delta_{ir}$ all pick up a minus sign.

\begin{figure}[h]
\centering
\includegraphics[width=1.9in]{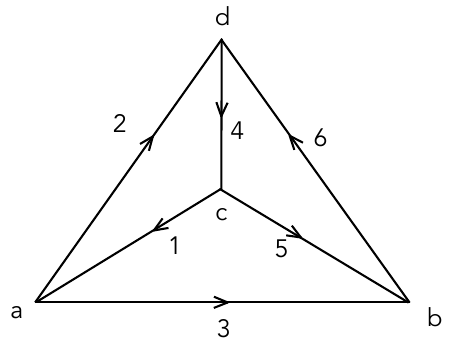}
\caption{ There is one tetrahedron diagram, which is related by symmetry to the tetrahedron diagram evaluated in Fig.~\ref{31loop}. }  \label{CubicVac1pA5reverse}
\end{figure}
We mentioned in the main body that, in addition to the tetrahedron diagram  in Fig.~\ref{31loop} that we evaluated, there is another diagram, which is  just Fig.~\ref{31loop} with the arrow on line $5$ reversed, as shown in Fig.~\ref{CubicVac1pA5reverse}. In our new notation, $5\rightarrow -5$. This means that its contribution is (\ref{329}) with $n_5 \rightarrow - n_5$, $\o_{p_5} \rightarrow - \o_{p_5}$, and the couplings $ \lam_{356}^* \lam_{145}$ replaced by $ -\lam_{635} \lam_{451}^*$.

\subsection*{Symmetries of the diagrams}
We will now show that the different contributions of a single tetrahedron diagram are  related by symmetry as well. 

\subsubsection*{Transformations of a tetrahedron} Let us look at the tetrahedron diagram studied in Sec.~\ref{cubicMain}, and shown again below in Fig.~\ref{teta6}(a), and consider all its symmetry transformation. There are $24$ in total, corresponding to the $4!$ possible positionings of the vertices. The symmetry transformations that we use are rotations of the tetrahedron as well as mirror reflections. For instance, in Fig.~\ref{teta6} we have shown the $6$ transformations that leave vertex $a$ fixed:  Fig.~\ref{teta6} (a) is the identity,  Fig.~\ref{teta6} (b) is a rotation about vertex a,  Fig.~\ref{teta6} (c) is a further rotation. Fig.~\ref{teta6} (d) is a mirror reflection of Fig.~\ref{teta6} (a) along the plane holding the $1$ line fixed and bisecting the face bordered by lines $2,3,6$.  Fig.~\ref{teta6} (e) is a mirror reflection of Fig.~\ref{teta6} (b) holding the $2$ line fixed, and Fig.~\ref{teta6} (f) is a mirror reflection of Fig.~\ref{teta6} (c) holding the $3$ line fixed. 
\begin{figure}[h]
\centering
\subfloat[]{
\includegraphics[width=1.6in]{CubicVac1pA.pdf}} \ \ \ 
\subfloat[]{
\includegraphics[width=1.6in]{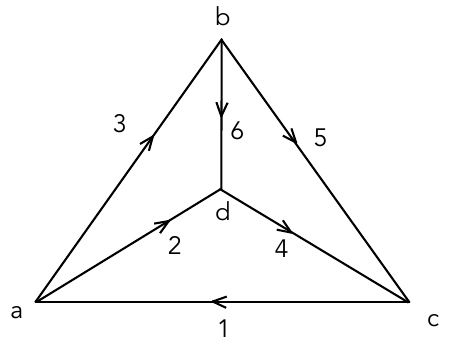}} \ \ \ 
\subfloat[]{
\includegraphics[width=1.6in]{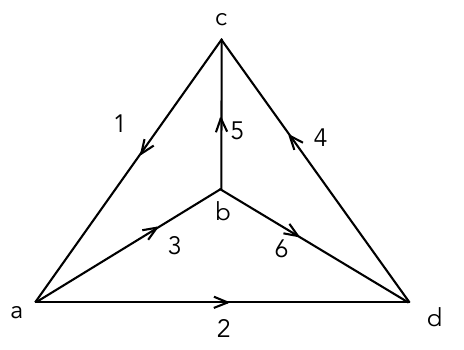}} \ \ \ 
\subfloat[]{
\includegraphics[width=1.6in]{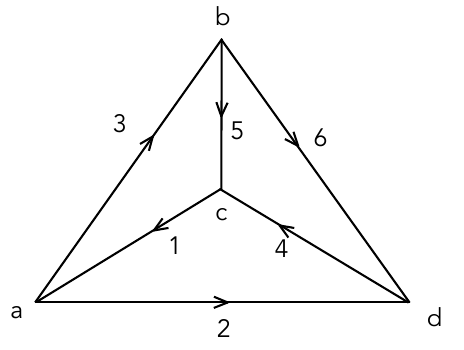}}\ \ \ 
\subfloat[]{
\includegraphics[width=1.6in]{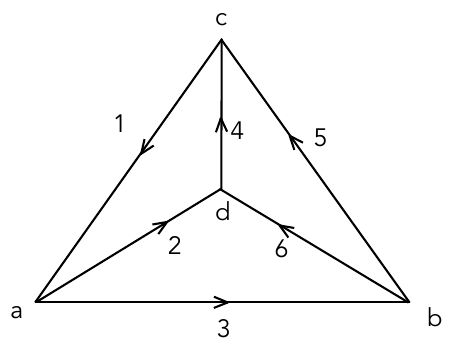}}\ \ \ 
\subfloat[]{
\includegraphics[width=1.6in]{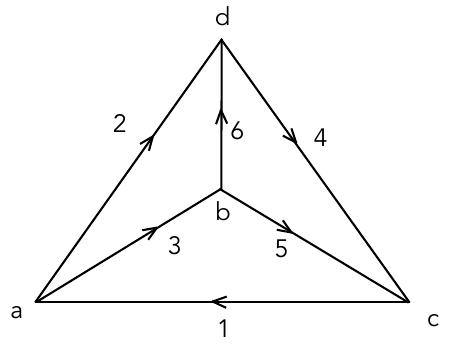}}
\caption{   The vertex orderings of the tetrahedra are: a) $a, b, c,d$;  b) $a, c,d, b$; c) $a,d,b,c$; d) $a,d,c,b$; e) $a,b,d,c$; f) $a,c,b,d$~. }  \label{teta6}
\end{figure}
Fig.~\ref{teta6}(a) has vertices labeled $a,b,c,d$ and edges labelled $1,2,3,4,5,6$. This is recorded as the first entry in our table below. We see that, for instance, Fig.~\ref{teta6}(e) has  vertices  $a,b,d,c$ and edges $-2,-1,3,-4,6,5$, where we are using a minus sign to denote the arrow running in the opposite direction. We have recorded this as the second entry in our table. The other entries are found in a similar manner. 

\begin{center}
\begin{tabular}{|c|c|}
\hline 
vertex ordering & edges \\
\hline
$a,b,c,d$ & $1,2,3,4,5,6$\\
\hline
$a,b,d,c$ & $-2,-1,3,-4,6,5$\\
\hline
$a,c,b,d$ & $-3,2,-1,-6,-5,-4$\\
\hline
$a,c,d,b$ & $-2,3,-1,6,-4,-5$\\
\hline
$a,d,b,c$ & $-3,-1,2,-5,-6,4$\\
\hline
$a,d,c,b$ & $1,3,2,5,4,-6$\\
\hline
$b,a,c,d$ & $-5,6,-3,4,-1,2$\\
\hline
$b,a,d,c$ & $-6,5,-3,-4,2,-1$\\
\hline
$b,c,a,d$ & $3,6,5,-2,1,-4$\\
\hline
$b,c,d,a$ & $-6,-3,5,2,-4,1$\\
\hline
$b,d,a,c$ & $3,5,6,1,-2,4$\\
\hline
$b,d,c,a$ & $-5,-3,6,-1,4,-2$\\
\hline
\end{tabular}
\quad \quad \quad
\begin{tabular}{|c|c|}
\hline 
vertex ordering & edges \\
\hline
$c,a,b,d$ & $5,-4,1,-6,3,2$\\
\hline
$c,a,d,b$ & $4,-5,1,6,2,3$\\
\hline
$c,b,a,d$ & $-1,-4,-5,-2,-3,6$\\
\hline
$c,b,d,a$ & $4,1,-5,2,6,-3$\\
\hline
$c,d,a,b$ & $-1,-5,-4,-3,-2,-6$\\
\hline
$c,d,b,a$ & $5,1,-4,3,-6,-2$\\
\hline
$d,a,b,c$ & $6,4,-2,-5,3,-1$\\
\hline
$d,a,c,b$ & $-4,-6,-2,5,-1,3$\\
\hline
$d,b,a,c$ & $2,4,-6,1,-3,5$\\
\hline
$d,b,c,a$ & $-4,-2,-6,-1,5,-3$\\
\hline
$d,c,a,b$ & $2,-6,4,-3,1,-5$\\
\hline
$d,c,b,a$ & $6,-2,4,3,-5,1$\\
\hline
\end{tabular}
\end{center}


\subsubsection*{Transforming terms in the kinetic equation}

Let us now show how we can use these symmetries to relate different terms in the kinetic equation. In particular, 
we start with one of terms, (\ref{Jrabcd1}), and define the right hand side of (\ref{Jrabcd1}) without the couplings to be $t(a,b,c,d)$,
\bea \nn
t(a,b,c,d)&=& \tilde t (a,b,c,d) \(\delta_{1r} {-}\delta_{2r} {-}\delta_{3r}\)~,\\
\tilde t(a,b,c,d)&=&G(\v e_{1;23}) G(\v e_{1; 256}) G(\v e_{4;26}) 
\frac{1}{n_3}\Big(\frac{1}{n_1}{-}\frac{1}{n_5}\Big)\Big(\frac{1}{n_2} {+}\frac{1}{n_6}{-}\frac{1}{n_4}\Big)\prod_{i=1}^6 n_i~. \label{B4}
\eea
We  view $t(a,b,c,d)$ as a function of the six edges, $t(1,2,3,4,5,6)$. The contribution to the kinetic equation of  the other  terms will still be given by $t$, but the arguments will contain some permutation of these six variables, and they may have minus signs. For instance, consider the term in which we have swap  vertices $c$ and $d$, which by our table corresponds to,
\bml
t(a,b,d,c) = t(-2,-1,3,-4,6,5)\\
= -\(\delta_{1r} {-}\delta_{2r} {-}\delta_{3r}\)G(\v e_{1;23}) G(\v e_{1; 256}) G(\v e_{1;45}) 
\frac{1}{n_3}\Big(\frac{-1}{n_2}{-}\frac{1}{n_6}\Big)\Big(\frac{-1}{n_1} {+}\frac{1}{n_5}{+}\frac{1}{n_4}\Big)\prod_{i=1}^6 n_i~.
\end{multline}
Under this permutation, the couplings transform as 
\be
 \lam_{123}\lam_{426}^* \lam_{356} \lam_{145}^* \rightarrow \lam_{-2 -1 3} \lam_{-4 -1 5}^* \lam_{3 6 5} \lam_{-2 -4 6}^* = \lam_{123}\lam_{426}^* \lam_{356} \lam_{145}^*~.
 \ee
We now collect all $6$ terms that have vertex $a$ first. We denote its contribution to the kinetic equation by, 
 \be \label{Ta}
 \sum_{1,\ldots, 6} \lam_{123}\lam_{426}^* \lam_{356} \lam_{145}^* \(\delta_{1r} {-}\delta_{2r} {-}\delta_{3r}\)T_a
 \ee
 \be \nn
 T_a = \tilde t(a,b,c,d) +\tilde t(a,b,d,c) -  \tilde t(a,c,b,d)- \tilde t(a,c,d,b)- \tilde t(a,d,b,c) - \tilde t(a,d,c,b)~.
 \ee
 The minus sign factors are due to the couplings picking up a minus sign. If we wish we can write this explicitly, 
 \bml \label{46}
T_a = G(\v e_{1;23}) \prod_{i=1}^6 n_i  
\Big\{ \frac{1}{n_1} G(\v e_{45;23})\Big[ G(\v e_{56;3}) \Big(\frac{1}{n_4}{-}\frac{1}{n_2}\Big)\Big(\frac{1}{n_5} {+}\frac{1}{n_6}{-}\frac{1}{n_3}\Big) + G(\v e_{4;26}) \Big(\frac{1}{n_5}{-}\frac{1}{n_3}\Big)\Big(\frac{1}{n_4} {-}\frac{1}{n_2}{-}\frac{1}{n_6}\Big) \Big] \\
 -\frac{1}{n_2} G(\v e_{16;34})\Big[ G(\v e_{56;3}) \Big(\frac{1}{n_1}{-}\frac{1}{n_4}\Big)\Big(\frac{1}{n_5} {+}\frac{1}{n_6}{-}\frac{1}{n_3}\Big) + G(\v e_{1;45}) \Big(\frac{1}{n_6}{-}\frac{1}{n_3}\Big)\Big(\frac{1}{n_1} {-}\frac{1}{n_4}{-}\frac{1}{n_5}\Big) \Big] \\
 -\frac{1}{n_3} G(\v e_{1;256})\Big[ G(\v e_{1;45}) \Big(\frac{-1}{n_2}{-}\frac{1}{n_6}\Big)\Big(\frac{1}{n_1} {-}\frac{1}{n_4}{-}\frac{1}{n_5}\Big) + G(\v e_{4;26}) \Big(\frac{1}{n_1}{-}\frac{1}{n_5}\Big)\Big(\frac{-1}{n_2} {+}\frac{1}{n_4}{-}\frac{1}{n_6}\Big) \Big] \Big\}~.
\end{multline}
This of course is just (\ref{329}). 
Now, the contribution of the terms with vertex $c$ first is, 
 \be -
  \sum_{1,\ldots, 6} \lam_{123}\lam_{426}^* \lam_{356} \lam_{145}^*\(\delta_{1r} {-}\delta_{4r} {-}\delta_{5r}\)T_c~,
 \ee
 where $T_c$ can be related to  $T_a$ by a  rotation of the tetrahedron that transforms vertex $a$ into vertex $c$. There are multiple ways of doing this, which differ by permutations of the other three vertices. Picking the rotation $\{a,b,c,d\} \rightarrow \{c,d,a,b\}$, and using the corresponding entry in the table, we have, 
\be
T_c= T_a(-1,-5,-4,-3,-2,-6)=T_a(1,5,4,3,2,6)^*~,
\ee
where in the second equality we  used  that flipping all the arrows corresponds to complex conjugating.  Since  under the change of variables $2, 3 \leftrightarrow 5,4$, the product of couplings transforms into its complex conjugate, $\lam_{123}\lam_{426}^* \lam_{356} \lam_{145}^* \rightarrow \lam_{123}^*\lam_{426} \lam_{356}^* \lam_{145}$,  the sum of the terms in which vertex $a$ is first and vertex $c$ is first is,
\be 
2i\, \text{Im} \sum_{1,\ldots, 6}\(\delta_{1r}{-}\delta_{2r}{-}\delta_{3r}\)  \lam_{123}\lam_{426}^* \lam_{356} \lam_{145}^*\, T_a(1,2,3,4,5,6)~,
 \ee
as was stated earlier. 
Next, we look at the contribution of terms in which vertex $b$ is first, 
 \bml \label{Tbsum}
-\sum_{1,\ldots, 6}\lam_{123}\lam_{426}^* \lam_{356} \lam_{145}^* (\delta_{3r} {-}\delta_{5r} {-}\delta_{6r}) T_b\\
T_b = \t t(b,a,c,d)+\t t(b,a,d,c) + \t t(b,c,a,d) + \t t(b,c,d,a) + \t t(b,d,a,c) + \t t(b,d,c,a)~.
\end{multline}
There is no need to compute $T_b$, because it can be related in a simple way to $T_a$: we simply rotate vertex $a$ into vertex $b$. There are multiple of doing this, which differ by permutations of the other three vertices. Picking the rotations $\{a,b,c,d\} \rightarrow \{b,d,a,c\}$, and using the corresponding entry in the table, we have, 
\be
T_b= T_a(3,5,6,1,-2,4)~.
\ee
Note the minus sign in front of (\ref{Tbsum}) was placed there to   account for minus sign acquired by the couplings under this transformation. 
Explicitly, $T_b$ is,
\bml \label{47}
T_b=G(\v e_{3;56}) \prod_{i=1}^6 n_i  
\Big\{ \frac{1}{n_3} G(\v e_{1;256})\Big[ G(\v e_{4;26}) \Big(\frac{1}{n_1}{-}\frac{1}{n_5}\Big)\Big(\frac{-1}{n_2} {-}\frac{1}{n_6}{+}\frac{1}{n_4}\Big) + G(\v e_{1;45}) \Big(\frac{1}{n_2}{+}\frac{1}{n_6}\Big)\Big(\frac{1}{n_4} {+}\frac{1}{n_5}{-}\frac{1}{n_1}\Big) \Big] \\
 -\frac{1}{n_5} G(\v e_{34;16})\Big[ G(\v e_{4;26}) \Big(\frac{1}{n_1}{-}\frac{1}{n_3}\Big)\Big(\frac{1}{n_2} {-}\frac{1}{n_4}{+}\frac{1}{n_6}\Big) + G(\v e_{23;1}) \Big(\frac{1}{n_6}{-}\frac{1}{n_4}\Big)\Big(\frac{1}{n_1} {-}\frac{1}{n_2}{-}\frac{1}{n_3}\Big) \Big] \\
 -\frac{1}{n_6} G(\v e_{23;45})\Big[ G(\v e_{1;45}) \Big(\frac{-1}{n_3}{-}\frac{1}{n_2}\Big)\Big(\frac{1}{n_4} {+}\frac{1}{n_5}{-}\frac{1}{n_1}\Big) + G(\v e_{23;1}) \Big(\frac{1}{n_4}{+}\frac{1}{n_5}\Big)\Big(\frac{1}{n_1} {-}\frac{1}{n_2}{-}\frac{1}{n_3}\Big) \Big] \Big\}~.
\end{multline}
Finally, the contribution of terms with vertex $d$ first is, 
 \be 
 \sum_{1,\ldots, 6} \lam_{123}\lam_{426}^* \lam_{356} \lam_{145}^*\(\delta_{4r} {-}\delta_{2r} {-}\delta_{6r}\)T_d~,
 \ee
where $T_d$ is related to $T_a$ by a rotation that  transforms vertex $a$ into vertex $d$. We pick the rotation $\{a,b,c,d\} \rightarrow \{d,b,c,a\}$, and using the corresponding entry in the table to get
\be
T_d= T_a(-4,-2,-6,-1,5,-3)=T_a(4,2,6,1,-5,3)^* = T_b(1,5,4,3,2,6)^*~.
\ee
To summarize we have shown that all contributions to the kinetic equation from the tetrahedron diagrams are symmetry transformations of one of the terms,  such as (\ref{B4}).

\section{Manipulating the kinetic equation} \label{apc}

In the main body of the text we  found that the contribution of the tetrahedron diagram to the kinetic equation is, 
\be \label{Tsum5}
\!\!\!\!\! 2i\, \text{Im} \sum_{1,\ldots, 6}\(\delta_{1r}{-}\delta_{2r}{-}\delta_{3r}\)\!\! \Big[  \lam_{123}\lam_{426}^* \lam_{356} \lam_{145}^*T_a(1,2,3,4,5,6) { -}  \lam_{123}\lam_{426}^* \lam_{635}^* \lam_{451}\ T_a(1,2,3,4,-5,6) \Big]
 \ee
 where $T_a$ was given in (\ref{46}) or equivalently the complex conjugate of (\ref{329}). The second term is the tetrahedron diagram in Fig.~\ref{31loop} with the arrow on line $5$ reversed, as discussed in Appendix~\ref{sec:sym}. 
 This form of the kinetic equation is acceptable, but the form of the answer that has a more clear physical interpretation is one  in which there are explicit delta functions. In particular, we write, 
 \be \label{osplit}
\frac{1}{\o_{p_1;p_2 p_3} {+}i\eps}  = \frac{1}{\o_{p_1;p_2p_3}} - i \pi \delta(\o_{p_1;p_2p_3})~,
 \ee
 where the first term comes with an implicit principal value. Using this, we may group terms based on the number of delta functions. This is what we partially do in this appendix. 
 
 One consistency check for our kinetic equation is that the thermal state should be stationary. In other words,  inserting $n_i = 1/\o_{p_i}$ into (\ref{Tsum5}) should give zero. This is not manifest for the piece of (\ref{Tsum5}) that has no delta functions. We will rewrite (\ref{Tsum5}) in a way that will make it manifest, by showing that there is no such piece. 

Another way of writing (\ref{Tsum5}) is as, 
 \ \be \label{Tsum}
\!\!\!\! \!\!\!\! \sum_{1,\ldots, 6}\! \lam_{123}\lam_{426}^* \lam_{356} \lam_{145}^*\Big[\!\! \(\delta_{1r} {-}\delta_{2r} {-}\delta_{3r}\)T_a{-}\(\delta_{3r} {-}\delta_{5r} {-}\delta_{6r}\) T_b{-}\(\delta_{1r} {-}\delta_{4r} {-}\delta_{5r}\)T_c{ +}\(\delta_{4r} {-}\delta_{2r} {-}\delta_{6r}\)T_d\Big]
 \ee
 where $T_b, T_c, T_d$ were defined in Appendix~\ref{sec:sym}. We may rewrite (\ref{Tsum}), grouping terms based on which $\delta_{ir}$ they come with, 
\be \label{Tsum2}
\!\! \sum_{1,\ldots, 6}\!\! \lam_{123}\lam_{426}^* \lam_{356} \lam_{145}^*\Big[ \delta_{1r}\(T_a {-} T_c\) {-}\delta_{2r}\(T_a {+}T_d\)  {-}\delta_{3r}\(T_a{ + }T_b\) + \delta_{4r}\(T_c {+} T_d\) + \delta_{5r} \(T_b {+} T_c\) + \delta_{6r}\(T_b{ -} T_d\) \!\Big]
 \ee
 If we use (\ref{osplit}) and look at the piece with no delta functions, we notice that each pair of parenthesis in (\ref{Tsum2}) vanishes, as it should.
  
Let us manipulate this part of the kinetic equation some more. Noting that under the change of variables $2, 3 \leftrightarrow 5,4$, the product of couplings transforms into its complex conjugate, $\lam_{123}\lam_{426}^* \lam_{356} \lam_{145}^* \rightarrow \lam_{123}^*\lam_{426} \lam_{356}^* \lam_{145}$, we may rewrite (\ref{Tsum2}) as,
\be \label{Tsum3}
2i\text{Im} \sum_{1,\ldots, 6} \lam_{123}\lam_{426}^* \lam_{356} \lam_{145}^*\Big[ \delta_{1r}T_a {-}\delta_{2r}\(T_a {+}T_d\)  {-}\delta_{3r}\(T_a{ + }T_b\) + \delta_{6r} T_b\Big]~.
 \ee
 Or, we may alternatively write it as, 
\be \label{Tsum4}
2i\text{Im} \sum_{1,\ldots, 6} \lam_{123}\lam_{426}^* \lam_{356} \lam_{145}^*\Big[ \(\delta_{1r}{-}\delta_{2r}{-}\delta_{3r}\) T_a  -  \(\delta_{3r}{-}\delta_{5r} {-}\delta_{6r}\)T_b \Big]~.
 \ee
If we wish, we can do a change of variables $(1,2,3,4,5,6) \rightarrow  (4,5,1,6,2,3)$ on the $T_b$ term, so as to rotate the $b$ vertex into the $a$ vertex to the greatest extent possible (meaning some of the arrows will be reversed; but we can't flip arrows under a change of variables. This gives back (\ref{Tsum5}). Note that from this perspective,  the second term in (\ref{Tsum5}) makes sense, because if we had done the transformation taking us from $a,b,c,d$ to $b,d,a,c$ then the edges would have gone from $1,2,3,4,5,6$ to $3,5,6,1,-2,4$ (see the table in Appendix.~\ref{sec:sym}), which is almost our transformation in the reverse direction, but with a minus sign for $5$.

 \subsection*{One delta function}
 Next, we look at the piece of \ref{Tsum5} which has one delta function. We will see that we can write this contribution to the kinetic equation entirely in terms of the following  two functions, 
\bea  \nn
D_{123}&=&\prod_{i=1}^6 n_i \!\(\frac{1}{n_1}{-}\frac{1}{n_2}{-}\frac{1}{n_3}\)\!\delta(\o_{p_1;p_2p_3})\\ \nn
&&\ \ \ \ \  \ \ \ \ \ \Big[\frac{-1}{n_4 n_5}\frac{1}{\o_{p_3;p_5p_6}}\frac{1}{\o_{p_4;p_2p_6}}+\frac{1}{n_4 n_6}\frac{1}{\o_{p_1;p_4p_5}}\frac{1}{\o_{p_3;p_5p_6}}+\frac{1}{n_5 n_6}\frac{1}{\o_{p_1;p_4p_5}}\frac{1}{\o_{p_4;p_2p_6}}\Big]~,\\ \label{D123}
D_{1256} &=&\prod_{i=1}^6 n_i \(\frac{1}{n_1}{+}\frac{1}{n_2}{-}\frac{1}{n_5}{-}\frac{1}{n_6}\) \delta(\o_{p_1p_2;p_5p_6})\frac{1}{n_1 n_6} \frac{1}{\o_{p_1;p_2p_3}}\frac{1}{\o_{p_3;p_5p_6}}~.
\eea
Like with $T_a$, these are functions of $(1,2,3,4,5,6)$. There are a few special vertex orderings we will want to consider. We write them, along with the corresponding edges. The two functions we just wrote are viewed as being at the vertex, 
$
A\equiv \{a,b,c,d\}=(1,2,3,4,5,6)
$.
We will also need, 
\bea
\!\!\!\!\!\!  \!\!\!\!\!\!  \!\!\!\!\!\!  A'&\equiv& \{a,d,c,b\}= (1,3,2,5,4,-6)~, \ \ \ B\equiv \{b,d,a,c\}=(3,5,6,1,-2,4)    \\ \nn
C&\equiv& \{c,d,a,b\}=(-1,-5,-4,-3,-2,-6)~, \ \ \ \ D\equiv \{d,b,c,a\}=(-4,-2,-6,-1,5,-3)~.
\eea
We find that the piece of $T_a$ that has terms with only one delta function is, 
\be \label{C10}
T_a = \[D_{123}(A) + D_{123}(B)- D_{123}(C) +D_{123}(D) \]+ \[(D_{1256}(A)-D_{1256}(A')+D_{1256}(B) \]~.
\ee
Next, note that $T_b = T_a(B)$, $T_c = T_a(C)$, $T_d = T_a(D)$, and so we find, 
\bea \nn
T_b &=& \[D_{123}(A) + D_{123}(B)+ D_{123}(C) -D_{123}(D) \]+ \[{-}D_{1256}(A){-}D_{1256}(A')+D_{1256}(B) \]\\ \nn
T_c &=& \[{-}D_{123}(A) + D_{123}(B)+ D_{123}(C) +D_{123}(D) \]+ \[{-}D_{1256}(A)+D_{1256}(A')+D_{1256}(B) \]\\
T_d &=& \[D_{123}(A) - D_{123}(B)+ D_{123}(C) +D_{123}(D) \]+ \[D_{1256}(A)+D_{1256}(A')+D_{1256}(B) \]  \label{C11}
\eea
Inserting into (\ref{Tsum2}), we get for the terms with $D_{123}$,
\bml
 2\sum_{1,\ldots, 6} \lam_{123}\lam_{426}^* \lam_{356} \lam_{145}^*\Big[ \delta_{1r}\[D_{123}(A) {-} D_{123}(C)\] {-}\delta_{2r}\[D_{123}(A) {+}D_{123}(D)\]  {-}\delta_{3r}\[D_{123}(A){ + }D_{123}(B)\] \\
 + \delta_{4r}\[D_{123}(C) {+} D_{123}(D)\] + \delta_{5r} \[D_{123}(B ){+} D_{123}(C)\] + \delta_{6r}\[D_{123}(B){ -}D_{123}(D)\] \Big]~.
 \end{multline}
 By the same manipulations as done in the beginning of this section, this is equal to 
   \bml  \label{C13}
4\sum_{1,\ldots, 6}\!\(\delta_{1r}{-}\delta_{2r}{-}\delta_{3r}\) \Big[\text{Re} (  \lam_{123}\lam_{426}^* \lam_{356} \lam_{145}^* )D_{123}(1,2,3,4,5,6) \\
 - \text{Re}( \lam_{123}\lam_{426}^* \lam_{635}^* \lam_{451})\ D_{123}(1,2,3,4,-5,6) \Big]~.
 \end{multline}
 Now, for the terms with $D_{1256}$, looking at their appearance  in (\ref{Tsum2}) upon inserting (\ref{C10}) and (\ref{C11}) we get, 
 \bml
 2\sum_{1,\ldots, 6} \lam_{123}\lam_{426}^* \lam_{356} \lam_{145}^*\Big[(\delta_{1r} {-}\delta_{2r}{-}\delta_{5r}{-}\delta_{6r}) D_{1256}(A) - (\delta_{2r} {+}\delta_{3r}{-}\delta_{4r}{-}\delta_{5r}) D_{1256}(B)\\
 - (\delta_{1r} {+}\delta_{6r}{-}\delta_{3r}{-}\delta_{4r}) D_{1256}(A')  \Big]
 \end{multline}
  Note also that since $
 D_{1256}(1,2,3,4,5,6)$ is invariant under $2,3\leftrightarrow 5,4$ we can replace the couplings with their real part, 
  \bml
 2\sum_{1,\ldots, 6}\text{Re}( \lam_{123}\lam_{426}^* \lam_{356} \lam_{145}^*)\Big[(\delta_{1r} {-}\delta_{2r}{-}\delta_{5r}{-}\delta_{6r}) D_{1256}(A) - (\delta_{2r} {+}\delta_{3r}{-}\delta_{4r}{-}\delta_{5r}) D_{1256}(B)\\
 - (\delta_{1r} {+}\delta_{6r}{-}\delta_{3r}{-}\delta_{4r}) D_{1256}(A')  \Big]~.
 \end{multline}
Finally, we  rotate the $B$ vertex and the $A'$ vertex to the extent possible: for the term with $D_{1256}(B)$ we send  $(1,2,3,4,5,6) \rightarrow  (4,5,1,6,2,3)$, like we did after (\ref{Tsum4}), and for the  term with $D_{1256}(A')$ we send  $(1,2,3,4,5,6) \rightarrow  (1,3,2,5,4,6)$ and then do a change of variables $5\leftrightarrow 6$. We get, 
   \bml \label{C16}
 2\sum_{1,\ldots, 6}\Big[\text{Re}( \lam_{123}\lam_{426}^* \lam_{356} \lam_{145}^*)(\delta_{1r} {-}\delta_{2r}{-}\delta_{5r}{-}\delta_{6r}) D_{1256}(1,2,3,4,5,6) -\(\delta_{1r} {+}\delta_{5r}{-}\delta_{2r}{-}\delta_{6r}\) \\
 \Big(  \text{Re}(\lam_{123}\lam_{426}^* \lam_{635}^* \lam_{451})D_{1256}(1,2,3,4,-5,6)+ \text{Re}( \lam_{123} \lam_{146}^* \lam_{635}^*\lam_{245}) D_{1256}(1,2,3,4,6,-5)\Big) \Big]
 \end{multline}
 To summarize: we have taken the part of the kinetic equation  due to the tetrahedron diagrams and extracted and simplified the piece that has one delta function. The result is the sum of (\ref{C13}) and (\ref{C16}). 

One can continue, and extract the piece of the kinetic equation that has a product of two delta functions, and the piece that has a product of three delta functions. As we go to higher order in the nonlinearity, the kinetic equation will have products of an increasing number of delta functions. It would be useful to have a general prescription for extracting the coefficients of these delta functions. This would be a kind of extension of the cutting rules in finite temperature field theory, see e.g.~\cite{Bedaque:1996af}. 

\section{Contact terms and lollipop diagrams}\label{technical}

In this appendix, we provide justification for disregarding the square of the absolute value of $\frac{\delta H_{\text{int}}}{\delta a_k}$ in the interaction Lagrangian \reef{Lint} (redundant interaction). Furthermore, we demonstrate that the existence of a non-zero expectation value for the complex field $a_p$ in the cubic theory \reef{cubic} does not affect our findings. We begin by addressing the redundant interaction.

Let us consider an expectation value of an operator $\mathcal{O}(a)$ constructed from the fields $a_k$ and $a^*_k$. The $m$th order correction can be expressed as follows
\be
\langle \mathcal{O}(a) \rangle^{(m)} =\Big\langle \mathcal{O}(a)  {(-1)^m\over m!} \( \int dt \sum_k\frac{1}{2 \g_k n_k} \Big[-i\sd_k a_k\frac{\delta H_{int}}{\delta a_k} + \text{c.c.}\Big] \)^m \Big\rangle + \ldots~, \quad
\sd_k=\d_t + i \o_k {+}\g_k ~.
\label{mth-order}
\ee
The first term on the right-hand side is obtained by expanding the path integral to the $m$th order in the interaction displayed in \reef{Lint}. The ellipsis represent terms involving the square of the absolute value of $\frac{\delta H_{\text{int}}}{\delta a_k}$, which we have omitted in \reef{Lint}. Now let us consider a specific set of terms that arises from pairing $\sd_k a_k$ with $\sd_k^*a_k^*$ in the first term of the above expression. This yields the following,
\bea
&&\langle \mathcal{O}(a) \rangle^{(m)} =2 \,{m(m-1)\over 2} \int dt_1 \int dt_2 \sum_k \frac{1}{4 \g_k^2 n_k^2} \big\langle \sd_k^*a_k^*(t_1) \, \sd_ka_k(t_2) \big\rangle  
 \\
&& \times \Big \langle \mathcal{O}(a) \, \frac{\delta H_{int}(t_1)}{\delta a^*_k} \,  \frac{\delta H_{int}(t_2) }{\delta a_k}  { (-1)^m\over m!} \( \int dt \sum_k\frac{1}{2 \g_k n_k} \Big[-i\sd_k a_k\frac{\delta H_{int}}{\delta a_k} + \text{c.c.}\Big] \)^{m-2} \Big\rangle +  \ldots~,
\nonumber
\eea
where the overall constant represents a combinatorial factor that takes into account the possible number of pairings. Substituting 
\be
 \big\langle \sd_k^*a_k^*(t_1) \, \sd_ka_k(t_2) \big\rangle = 2\g_k n_k \delta(t_1-t_2) ~,
\ee
gives
\bea
&&\langle \mathcal{O}(a) \rangle^{(m)} =
 \\
&& \times \Big \langle \mathcal{O}(a)  \int dt_1  \sum_k \frac{1}{2 \g_k n_k}    \bigg|\frac{\delta H_{int}(t_1)}{\delta a^*_k}\bigg|^2 \,   { (-1)^{m-2}\over (m-2)!} \( \int dt \sum_k\frac{1}{2 \g_k n_k} \Big[-i\sd_k a_k\frac{\delta H_{int}}{\delta a_k} + \text{c.c.}\Big] \)^{m-2} \Big\rangle +  \ldots~.
\nonumber
\eea
Up to a sign, this expression is equal to the $m$th order term obtained by using $m-2$ interactions of the form shown in \reef{Lint} and one interaction involving the square of the absolute value of $\frac{\delta H_{\text{int}}}{\delta a_k}$. Hence, these terms mutually cancel each other. Extending this conclusion to any number of pairings between $\sd_k a_k$ and $\sd_k^*a_k^*$ in \reef{mth-order}, we can disregard the square of the absolute value of $\frac{\delta H_{\text{int}}}{\delta a_k}$ in the interaction Lagrangian $L_{\text{int}}$, as well as all diagrams resulting from pairing $\sd_k a_k$ with $\sd_k^*a_k^*$.

\begin{figure}[t]
\centering{\includegraphics[width=1.8in]{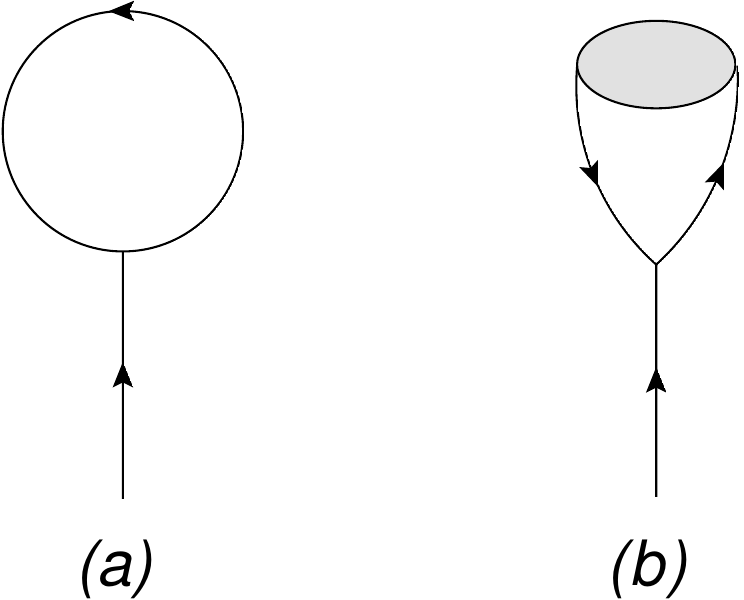}}
\caption{  (a) A one-loop diagram contributing to $\langle a_p\rangle$ (b) Higher order corrections to $\langle a_p\rangle$. The gray blob represents corrections to the propagator of $\langle a^*_p(t) a_p(0) \rangle$.}  
\label{lollipop}
\end{figure}

Next, we show how to get rid of the expectation value of the complex field $a_p$ in the cubic theory \reef{cubic}. To leading order $\langle a_p \rangle$ is given by the lollipop diagram in Fig.~\ref{lollipop}(a). This diagram can be  evaluated using the Feynman rules in the main body of the text,
\be
 \langle a_p \rangle= {-1\over \o_p - i \gamma_p }  \sum_k \lambda^*_{kpk} n_k + \mathcal{O}(\lambda^3)~.
 \label{backgrnd}
\ee
Therefore, considering the momentum-conserving delta function in $\lambda_{p_1p_2p_3}$, we find that $\langle a_p \rangle$ is proportional to the Dirac delta function $\delta(\vec p)$. It represents a shift from the zero value of the homogeneous background that defines the lowest energy state of the waves. The objective of our work is to study excitations around a constant background. To isolate degrees of freedom associated with excitations, we introduce the following {\it canonical} change of variables.
\be
  a_p \to a_p + b_0 \delta(\vec p)~, \quad  \da_p \to    a^\dagger_p + b^*_0 \delta(\vec p) ~,
\ee
where the complex constant $b_0$ is defined by $\langle a_p \rangle = b_0 \delta(\vec p)$; it can be calculated by evaluating the diagrams in Fig.~\ref{lollipop}(b). By definition, the expectation value of the shifted fields vanishes, and the Hamiltonian \reef{cubic} takes the form
\begin{eqnarray}
H &\to&   \sum_p \o_p\da_p a_p + \frac{1}{2}\sum_{p_i}\( \lam_{p_1 p_2 p_3} \da_{p_1} a_{p_2} a_{p_3} + \lam^*_{p_1 p_2 p_3} a_{p_1} \da_{p_2} \da_{p_3}\) + J_0\da_0 + J^*_0 a_0
\nonumber \\
&+& \frac{1}{2}\sum_{p_1,p_2}\( b_0^* \lam_{0 p_1 p_2}  a_{p_1} a_{p_2}  + b_0 \lam^*_{0 p_1 p_2}  \da_{p_1} \da_{p_2}+ 2 b_0 \lam_{p_1 p_2 0}  \da_{p_1}  a_{p_2} 
+ 2 b_0^*  \lam^*_{p_1 p_2 0} a_{p_1} \da_{p_2}  \) ~,
\end{eqnarray}
where $J_0=b_0\o_0 + |b_0|^2\sum_p \lam^*_{00p} +  {1\over 2} b_0^2\sum_p \lam_{p00} $, and we have dropped the field independent constants (we have renormalized the cosmological constant). Due to the momentum-conserving delta function in $\lambda_{p_1p_2p_3}$, one can cancel the last two terms in the second line by suitably redefining $\omega_p$ (frequency renormalization),
\be
 \omega_p \to \omega_p - \sum_{p_2} \(  b_0 \lam_{p p_2 0}  +  b_0^*  \lam^*_{p p_2 0}   \) ~.
\ee
Thus, we obtain
\begin{eqnarray}
H &\to&   \sum_p \o_p\da_p a_p + \frac{1}{2}\sum_{p_1,p_2}\( b_0^* \lam_{0 p_1 p_2}  a_{p_1} a_{p_2}  + b_0 \lam^*_{0 p_1 p_2}  \da_{p_1} \da_{p_2}  \)
\nonumber \\
&+&  \frac{1}{2}\sum_{p_i}\( \lam_{p_1 p_2 p_3} \da_{p_1} a_{p_2} a_{p_3} + \lam^*_{p_1 p_2 p_3} a_{p_1} \da_{p_2} \da_{p_3}\) + J_0\da_0 + J^*_0 a_0 ~.
\end{eqnarray}
To simplify this expression further, we employ a redefinition of the frequency as $\omega_p\to \mathcal{N}_p \, \omega_p$, along with a corresponding field redefinition given by 
\bea
 a_p &\to& \mathcal{N}^{-1/2}_p\Big(a_p -  \frac{1}{2 \o_p}\sum_{p_2} b_0 \lam^*_{0 p p_2} \da_{p_2} \Big) ~, \quad \mathcal{N}_p=1-{|b_0|^2\over 4\o_p^2}\sum_k |\lam_{0pk}|^2~,
 \nonumber \\
 \da_p &\to& \mathcal{N}^{-1/2}_p\Big( \da_p -  \frac{1}{2 \o_p}\sum_{p_2} b^*_0 \lam_{0 p p_2} a_{p_2} \Big)  ~.
 \label{redef} 
\eea 
This transformation is canonical. As a result, up to linear terms in the fields, the Hamiltonian takes the original form \reef{cubic}
\begin{eqnarray}
H =\sum_p \tilde \o_p\da_p a_p 
+  \frac{1}{2}\sum_{p_i}\( \tilde \lam_{p_1 p_2 p_3} \da_{p_1} a_{p_2} a_{p_3} + \tilde \lam^*_{p_1 p_2 p_3} a_{p_1} \da_{p_2} \da_{p_3}\) + \tilde J_0\da_0 +  \tilde J^*_0 a_0  ~,
\end{eqnarray}
where quantities with tilde absorb various terms which emerge due to the field redefinition \reef{redef}.\footnote{It should be noted that cubic interactions of the form $a_{p_2}a_{p_2}a_{p_3}$ and $\da_{p_1}\da_{p_2}\da_{p_3}$ can also be eliminated through an appropriate field redefinition \cite{Falkovich}. Consequently, we have omitted these terms in the final expression.} By construction, the last two terms guarantee that $\langle a_0 \rangle=0$.

We have shown that the expectation value of $a_p$ in the cubic theory can be ignored by appropriately redefining the frequencies, couplings, and fields. To achieve this, one introduces a linear term of the form $ J_0\da_0 +   J^*_0 a_0$ in the Hamiltonian \reef{cubic}, where the unspecified complex constant $J_0$ is appropriately tuned such that $\langle a_p \rangle=0$ to all orders of the perturbative expansion. In other words, we can safely disregard the lollipop diagrams in Fig.~\ref{lollipop}. The value of $\langle a_p \rangle$ is determined by minimizing the complete effective potential of the model, whereas the Hamiltonian \reef{cubic} describes only excitations.

\end{document}